\documentclass[pageno]{jpaper}

\usepackage{def}
\usepackage{package}

\usepackage[normalem]{ulem}

\newcommand{\memmove}{\texttt{memmove()}\xspace}
\newcommand{\memcpy}{\texttt{memcpy()}\xspace}
\newcommand{\memcpys}{\texttt{memcpy()}s\xspace}
\newcommand{\memcmp}{\texttt{memcmp()}\xspace}
\newcommand{\memset}{\texttt{memset()}\xspace}
\iftrue

\newcommand{\ipoom}[1]{{\color{blue}[\textbf{\sc ipoom}: \textit{#1}]}}
\newcommand{\reese}[1]{{\color{teal}[\textbf{\sc reese}: \textit{#1}]}}

\else
\newcommand{\nskim}[1]{}
\newcommand{\ipoom}[1]{}
\newcommand{\reese}[1]{}
\newcommand{\yifan}[1]{}

\fi

\newcommand\Tstrut{\rule{0pt}{2.5ex}}       	
\newcommand\Bstrut{\rule[-1.0ex]{0pt}{0pt}} 	
\newcommand{\TBstrut}{\Tstrut\Bstrut}			 
\newcolumntype{C}[1]{>{\centering\arraybackslash}p{#1}}
\def\accel{DSA\xspace}
\def\fig{Fig.\xspace}

\newcommand{\highlight}[1]{\fontsize{10.5}{11}\texttt{#1}}

\usepackage[normalem]{ulem}

\begin{document}

\title{A Quantitative Analysis and Guidelines of Data Streaming Accelerator \\ in Modern Intel\textsuperscript{\textregistered} Xeon\textsuperscript{\textregistered} Scalable Processors}

\date{}
 \author{
 {\rm Reese Kuper}\\
 \emph{UIUC} \\ rkuper2@illinois.edu
 \and
  {\rm Ipoom Jeong}\\
 \emph{UIUC} \\ ipoom@illinois.edu
 \and
 {\rm Yifan Yuan}\\
 \emph{Intel Labs} \\yifan.yuan@intel.com
 \and
 {\rm Jiayu Hu}\\
 \emph{Intel Corporation} \\jiayu.hu@intel.com
 \and
 {\rm Ren Wang}\\
 \emph{Intel Labs} \\ ren.wang@intel.com
 \and 
  {\rm Narayan Ranganathan}\\
 \emph{Intel Labs} \\narayan.ranganathan@intel.com
 \and 
 {\rm Nam Sung Kim}\\
 \emph{UIUC} \\nskim@illinois.edu
 } 
\maketitle

\thispagestyle{empty}

\begin{abstract}
\footnote{This work has been accepted by a conference. The authoritative version of this
work will appear in the Proceedings of the ACM Architectural Support for Programming Languages and Operating Systems (ASPLOS), 2024.}
As semiconductor power density is no longer constant with the technology process scaling down, modern CPUs are integrating capable data accelerators on chip, aiming to improve performance and efficiency for a wide range of applications and usages. 
One such accelerator is the Intel\textsuperscript{\textregistered} Data Streaming Accelerator (\accel) introduced in Intel\textsuperscript{\textregistered} 4th Generation Xeon\textsuperscript{\textregistered} Scalable CPUs (Sapphire Rapids).
\accel targets data movement operations in memory that are common sources of overhead in datacenter workloads and infrastructure.
In addition, it becomes much more versatile by supporting a wider range of operations on streaming data, such as CRC32 calculations, delta record creation/merging, and data integrity field (DIF) operations.
This paper sets out to introduce the latest features supported by \accel, deep-dive into its versatility, and analyze its throughput benefits through a comprehensive evaluation. 
Along with the analysis of its characteristics, and the rich software ecosystem of \accel, we summarize several insights and guidelines for the programmer to make the most out of \accel, and use an in-depth case study of DPDK Vhost to demonstrate how these guidelines benefit a real application.
\end{abstract}

\section{Introduction}\label{sec:introduction}
The demise of Dennard scaling~\cite{darksilicon:isca:2011} has given rise to the wide adoption of accelerators for common application domains, such as GPUs, NPUs, and more recently, DPUs~\cite{TPU,yazdanbakhsh2021evaluation,masterofnone:isca:2019,dadiannao:micro:2014,diannao:asplos:2014,streamdf:isca:2017,dnnweaver:micro:2016,tabla:hpca:2016,scalable:isca:2015,graphicionado:micro:2016,flexlearn:micro:2019,towardsgen:micro:2019,extensor:micro:2019,eyeriss:isca:2016,proto:micro:2021}. 
The rise of specialization also yields a trend towards core integrated with diverse on-chip accelerators such as Gaussian \& Neural Accelerator in Intel Core processors~\cite{intelgna31:online} and Active Messaging Engines in IBM Power10 processors~\cite{ibm-dma}. 

Datacenter servers often run not only diverse latency-sensitive applications, but also common software components such as data movement (e.g., \memcpy and \memmove), hashing, and compression.
The CPU cycles spent on running these common software components are known as \emph{datacenter tax}, since they should instead be used to run applications~\cite{Kanev:2015:PWC:2749469.2750392}.
Besides, these data movement/transform operations often compete with other co-running applications for the shared CPU resources, such as on-chip caches, and thus often notably impact application service latency.
For these reasons, efficient offload of such operations to optimized hardware engines is a promising approach in datacenter system-on-chips~\cite{Kanev:2015:PWC:2749469.2750392}.

To offload such memory operations from the CPU, Intel introduced I/O Acceleration Technology (I/OAT) to server-class Xeon\textsuperscript{\textregistered} 5100 series processors in 2006~\cite{18642woo73:online}. 
I/OAT is comprised of a DMA engine in the chipset that is recognized, programmed, and operated in the same way as DMA engines in PCIe-connected devices.
Generational improvements to I/OAT evolved into the Crystal Beach DMA (CBDMA) engine with increased throughput and data coordination. 
Despite having support from some software stacks, broader adoption by application stacks has proven harder due to the restricted programming interface and higher offload costs.

In response to the growing need for offloading more diverse, simple but repetitive memory-intensive operations, Intel recently introduced the Data Streaming Accelerator (\accel) to Intel\textsuperscript{\textregistered} 4th generation Xeon\textsuperscript{\textregistered} Scalable Processors based on the Sapphire Rapids architecture~\cite{Introduc99:online}.
Compared to DMA engines from previous generations, \accel has been improved significantly in various aspects, including substantially lower offload overhead, better efficiency in terms of throughput and energy consumption, diversity of supported operations, and a set of unique features which make it a more desirable and attractive solution for accelerating memory-related operations and transformations.

In this paper, we first investigate and quantify the benefits of \accel from various aspects by using a set of microbenchmarks. 
Subsequently, we demonstrate the software enablement and ecosystem of \accel in modern datacenter infrastructure and application domains.
We also present the real applications that gain performance improvement by adopting \accel.
Based on the performance analysis and software enablement experience, we provide some guidelines to make the best use of \accel features. 
Using DPDK Vhost as a tangible example, we also demonstrate how \accel usage can be adopted and optimized in a real application in detail~\cite{dpdk-vhost-lib:online}.
Our comprehensive analysis demonstrates that \accel has great potential for accelerating a number of widely used operations on streaming data for higher throughput and lower tail-latency, while preventing performance-critical on-chip shared cache from being polluted by such streaming data. 

\begin{figure*}[!t]
    \centerline{\includegraphics[width=0.95\textwidth]{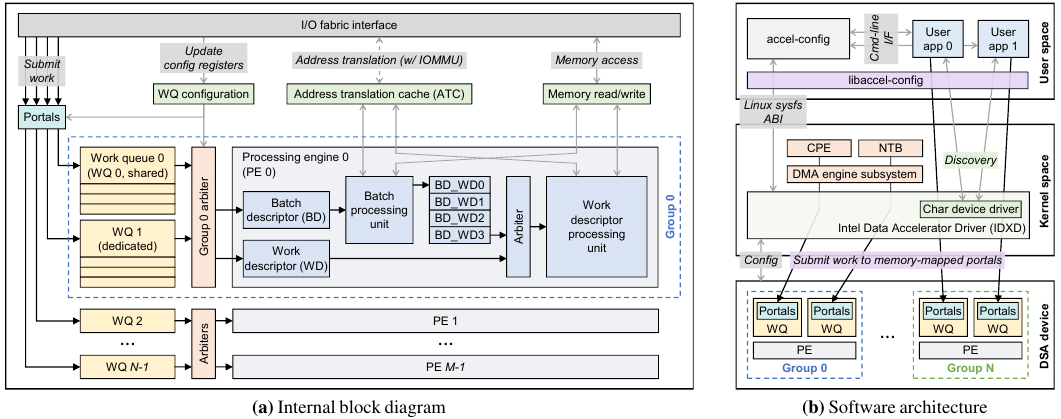}}
    \vspace{-4pt}
    \caption{Architectural overview of \accel. Job descriptors are directly submitted to memory-mapped portals in each device. IOMMU allows in-device address translation, and thus memory pinning is not required.}
    \label{fig:dsa_overview}
    \vspace{-6pt}
\end{figure*}

\section{Background and Motivation}\label{sec:background}

In modern server systems, there are various repetitive memory-intensive routines frequently deployed and invoked for several purposes~\cite{Kanev:2015:PWC:2749469.2750392}. 
One example is memory copying, which is ubiquitous in the scenarios of network packet processing and file I/O accesses. 
Other examples include \memcmp for comparing different regions of memory, \memset for assigning a value to a designated memory region, delta-record for tracking the difference between two memory regions, and data integrity checks like CRC checksums. 
Although some of them have been optimized in software or through new instructions, they still exhibit a significant cost on CPU core utilization in datacenters~\cite{Kanev:2015:PWC:2749469.2750392}.

To address the overheads of memory operations, Intel introduced QuickData Technology, a component of its I/OAT set~\cite{IntelIO3:online,WhitePap2:online}, that facilitates the \memcpy operations by the chipset instead of the cores~\cite{4629228,4228207,vaidyanathan2007benefits,goglin2008improving}.  
Evolving from the chipset, this technology has been implemented as an on-chip unit (i.e., CBDMA) in the past few generations of Intel Xeon Scalable Processors. 
A CPU core can offload memory copy operations to CBDMA in an asynchronous fashion by submitting a request descriptor to CBDMA and waiting for an interrupt upon completion.
As a result, the dedicated memory copy engine effectively saves CPU cycles. 
Despite this, CBDMA has shown limited use in general applications due to a small feature set, high offloading cost, and restrictions on its programmability, such as the requirement for memory pinning~\cite{quickdat48:online}.
Hence, a new hardware engine with better performance, more versatile functionalities, and more user-friendly programming model is highly desired. 

\section{Intel Data Streaming Accelerator}\label{sec:data-streaming-accelerator}

\subsection{Overview}\label{subsec:dsa_overview}

\accel is a high-performance data copy and transformation accelerator integrated in Intel's latest Sapphire Rapids Xeon CPUs as a successor of CBDMA~\cite{Introduc99:online}. It provides not only better processing efficiency and practicality, but also greater versatility over CBDMA. 
\accel is equipped with hardware components to efficiently process work descriptors submitted by one or more users (\S~\ref{subsec:dsa_hw_arch}). Through the support of shared virtual memory (SVM), these work descriptors can be submitted by user applications directly to \accel via memory-mapped I/O (MMIO) registers.
Users are also allowed to directly configure the underlying computational resources based on their needs, which is enabled by the user interfaces provided by a set of dedicated drivers (\S~\ref{subsec:dsa_sw_arch}). 
In addition, \accel provides better functionality by means of newly supported operations, and therefore it is expected that more user/kernel processes can take advantage of what \accel offers. 
Several optimization features are also introduced in \accel, aiming to make the best use of the accelerator's available resources (\S~\ref{subsec:dsa_key_features}).

\begin{table*}[t!]
\caption{Data Streaming Operations Supported by \accel}
\label{tab:dsa-operation}
\vspace{-10pt}
\begin{center}
\resizebox{0.9\textwidth}{!}{%
\begin{tabular}{|c|c|c|}
\hline
\textbf{Type} & \textbf{Operation} & \textbf{Description} \TBstrut\\
\hline
\multirow{5}{*}{Move} & Memory Copy & Copy data from a source address to a destination address \Tstrut  \\
 & Dualcast & Transfer data from a source address to two separate destination addresses  \\
 & CRC Generation & Create a CRC checksum on the transferred source data  \\
 & Data Integrity Field & Check, insert, strip, or update DIFs for each 512/520/4096/4104-byte block of source data \Tstrut\\
 & (DIF) Operations & transferred from a source address to a destination address \Bstrut  \\
\hline
Fill & Memory Fill & Sequentially fill the designated memory region with a 8/16-byte fixed pattern\TBstrut \\
\hline
\multirow{4}{*}{Compare} & Memory Compare & Compare two source buffers against each other and return whether they are identical\Tstrut  \\
 & Compare Pattern & Compare a source buffer with an 8-byte pattern and return whether the entire region matches the pattern  \\
 & Create Delta Record & Creates a delta record containing the differences between two source buffers (i.e., original/modified buffers)\\
 & Apply Delta Record & Merges a delta record to the original buffer to create the copy of the modified buffer to a destination address\Bstrut \\
\hline 
Flush & Cache Flush & Evict all cache lines within a given address range from the cache hierarchy\TBstrut \\
\hline 
\multicolumn{3}{l}{\textit{* Note: Create and Apply Delta Record and Cache Flush are not analyzed due to being more niche and specialty operations.}} \Tstrut\\
\end{tabular}
}
\label{tab2}
\end{center}
\vspace{-12pt}
\end{table*}

\subsection{Hardware Architecture}\label{subsec:dsa_hw_arch}

\fig~\ref{fig:dsa_overview}(a) illustrates the high-level hardware organization of \accel. It consists of a set of interfaces to communicate with the host and other system components (e.g., main memory), work queues (WQs) that hold the submitted work descriptors, processing engines (PEs) that fetch the descriptors and conduct requested work, and arbiters for quality of service (QoS) and fairness control. The basic operational unit of \accel is a group, which can be flexibly configured with any number of WQs and PEs. The WQs in a group accept descriptors from software and the PEs in the group process the descriptors from the WQs. In the figure, \texttt{Group 0} is configured to have two WQs and one PE. The performance impact of group configuration is further discussed in \S~\ref{subsec:configurable_groups}.

As an on-chip accelerator, each \accel device is exposed to the host as a single root complex integrated endpoint (RCiEP) and fully compatible with the standard PCIe configuration mechanism. 
Software offloads work by submitting 64-byte work descriptors to \accel via device MMIO registers (called ``portals''). The work descriptor includes the key information and settings of the operation, such as operation type, completion record address, ordering of the device memory write, batching, etc.
Work descriptors submitted to a portal are placed in the WQ corresponding to the portal address. The WQs are on-device storage configured by user-level utility before enabling the device. Each WQ operates in one of two modes: (1) dedicated WQ (DWQ), which are intended or a single client to submit work to, and (2) shared WQ (SWQ), which can be shared by multiple software clients without inter-thread synchronization.

A descriptor at the head of a WQ is eligible to be dispatched to any free PE within a group. This is performed by the group arbiter, considering the relative priority of the WQs inside a group for QoS, while preventing starvation. Then, the work descriptor processing unit conducts the operation specified in a descriptor by: (1) reading source data from the memory, (2) performing operation, (3) updating destination buffer in the memory, and (4) writing completion record and generating interrupted if requested. See \S~\ref{subsec:dsa_key_features} for details on batch descriptor processing and QoS control. 

Meanwhile, the address translations for the completion record, source, and destination buffers are performed by interacting with the on-device address translation cache (ATC) that interacts with the IOMMU on the SoC --- a key difference from previous generations. This enables support of coherent shared memory between \accel and cores --- they can access shared data in CPU virtual address space and thereby eliminate the need for applications to pin memory.

\subsection{Software Architecture and Interface}\label{subsec:dsa_sw_arch}

\fig~\ref{fig:dsa_overview}(b) represents the software architecture of \accel. 
The key software is Intel Data Accelerator Driver (IDXD) -- a kernel mode driver for device initialization and management. 
For the control path, IDXD provides functionalities for \accel discovery, initialization, and configuration.
Applications can leverage the user-space \texttt{libaccel-config} API library for such control operations. 
For the data path, to provide low-latency access to the \accel instance for the applications, IDXD exposes the MMIO portals as a \texttt{char} device via \texttt{mmap}. Note that IDXD also enables in-kernel usage of \accel (e.g., clear page engine \texttt{CPE} and non-transparent bridge \texttt{NTB})~\cite{ntb-driver}.

As discussed before, \accel is invoked via the submission of a work descriptor and it generates a corresponding completion record when the offloaded task is completed. To improve communication efficiency between the cores and \accel, several new x86 instructions for low-latency work-dispatch and synchronization are supported on the CPU cores.

For low-latency work submission to dedicated work queues (DWQ), the \texttt{MOVDIR64B} instruction is introduced.
For the submission to a SWQ, the \texttt{ENQCMD} and \texttt{ENQCMDS} instructions are used instead to avoid multi-producer contention, removing the need for software to maintain locks to perform descriptor submissions. These instructions atomically submit one descriptor to the WQ in a non-posted manner and a status value is returned showing whether the descriptor has been successfully accepted by \accel. If unsuccessful, the submitter can re-try the descriptor submission if needed. 

For the core to get the results from \accel asynchronously while avoiding high cycle and power consumption for spin-polling, the \texttt{MONITOR/MWAIT} instructions have been extended to user space (i.e., \texttt{UMONITOR/UMWAIT}) so that user space software can now register and monitor a memory address, while remaining in an optimized wait state until the value is changed (e.g., completion record updated by \accel).

\subsection{Supported Operations and Unique Key Features}\label{subsec:dsa_key_features}

\accel supports a rich set of operations, as listed in Table~\ref{tab:dsa-operation}. More details can be found in the official specification~\cite{dsa-spec}.
In this section, we describe unique key \textbf{\underline{F}}eatures provided by \accel compared to its predecessors.

{\noindent \textbf{F1: Virtualization and multi-application support.}}
\accel is able to operate on either virtual addresses or physical addresses, such as for specifying the locations of the source and destination buffers. As it has the ability to access the virtual address space of the host applications (i.e., SVM), task offloading does not require memory pinning of source and destination buffers. Instead, process address space ID (PASID) is used for looking up addresses in the ATC, sending requests for address translation to IOMMU (i.e., page table walk), and handling page faults via communication with the OS. This feature enables multiple applications (processes) to use \accel simultaneously and independently. \accel also supports I/O virtualization (Intel Scalable IOV~\cite{Introduc28:online}), enabling it to be shared by multiple VMs in secure and isolated environments.

\noindent \textbf{F2: Batch descriptor processing.}
\accel supports the processing of a batch descriptor that delivers the starting address of an array of work descriptors and the number of them in the batch to reduce the overhead of submitting multiple work descriptors one by one. Being prepared and submitted in the same way as work descriptors, a batch descriptor is handled by the batch processing engine: fetching a set of work descriptors from memory at once, and subsequently processing them in the work descriptor processing unit. In addition, a batch-granular completion record is generated, which alleviates the overhead of software to examine the completion records of all batched descriptors. As a result, batch description processing significantly improves data throughput over running these tasks individually on \accel, as well as on a core. The performance impact of batch processing is explored in great detail in \S~\ref{subsec:throughput_and_latency}.

\noindent \textbf{F3: QoS control.}
A number of QoS features are provided for the finer-grained control over the latency- and bandwidth-sensitive components inside \accel. Priority values can be set to WQs, adjusting the frequency at which descriptors are dispatched to the PEs. Co-running applications that make use of multiple WQs per group can set an appropriate priority to ensure sufficient resources for the other running processes that access the same group.

\accel includes a set of features to assist working with memory types and tiers that vary in bandwidth and latency characteristics. External to the accelerator, \accel allows access to the underlying PCIe fabric, enabling support for traffic classes and virtual channels. Internally, read buffers are used to hide memory access latency and maximize engine throughput. These buffers, much like the other components in \accel, can be configured. Decreasing the number of read buffers for a PE may affect its achievable bandwidth, but it also frees read buffers that can then be allocated to other engines. Additional configuration options are also available -- such as the maximum batch size, WQ size, and transfer size -- which provide users with greater control over expected offloading latency.

\begin{table}[!t]
\centering
\caption{Evaluated System Configurations}
\vspace{-4pt}
\scriptsize
\resizebox{\columnwidth}{!}{%
\begin{tabular}{ccc}
\hline
\textbf{Generation} &  \textbf{Ice Lake (ICX)} & \textbf{Sapphire Rapids (SPR)}\TBstrut \\
\hline
Number of cores                             & 40            & 56                   \\
L1I/L1D/L2 (KB)                                 & 32 / 48  / 1280     & 32 / 48 / 2048             \\
Shared LLC (MB)                             & 57            & 105                  \Bstrut  \\
\hline
Memory (GB)                                 & Six DDR4 Channels    & Eight DDR5 Channels           \TBstrut \\
\hline
\multirow{2}{*}{DMA engine}                 & CBDMA w/      & \accel w/ 8 WQs,     \Tstrut \\
                                            & 16 channels   & 4 engines            \Bstrut \\
\hline
\end{tabular}
}
\label{tab1}
\vspace{-4pt}
\end{table}

\section{\accel Performance Analysis}\label{sec:result_and_analysis}

\subsection{Experimental Methodology}\label{sec:methodology}

In our experiments, we evaluate two DMA engines in Intel 4th gen Xeon CPU (SPR) and its predecessor (ICX), as shown in Table~\ref{tab1}. Although CBDMA and \accel have different hardware architectures and software interfaces, we try to use (logically and/or physically) the same resources across different DMA engines for a fair comparison (e.g., a single channel for CBDMA and a single PE for \accel). In the same vein, the cache control feature described in \S~\ref{subsec:interation_with_cache_mem} is disabled for \accel. Instead, we highlight the effectiveness of other basic functionalities. 
We use Linux kernel version 5.15 when measuring latency, throughput, and cache pollution effects.

For benchmarks, we use a set of user-space microbenchmarks, \texttt{dsa-perf-micros} ~\cite{dsa-perf-micros:online}, to measure the performance of \accel with different group configurations, job descriptions, and synchronization modes. When used asynchronously, \accel is used with a queue depth of 32 unless otherwise stated. Each test runs over one thousand iterations and input data is initialized in the main memory at the beginning of every test case.
For the software baseline running on the CPU core, we select highly optimized software libraries (e.g., \texttt{glibc}'s \texttt{memcpy}, and ISA-L~\cite{isa-l} for CRC32 calculation) for fair comparison.
    Latency breakdown measurements were measured by modifying \texttt{dsa\_test} found in \accel's user utility, \texttt{accel-config}, from the open-source \texttt{idxd-config} repository~\cite{idxd-config:online}. Specifically, we set the system timers around the regions of code used for allocating, preparing, submitting, and waiting for the descriptor to finish processing. For comparisons against the baseline CPU, we made a software counterpart of each operation and timed them accordingly. To ensure consistency within our testing, we flushed the descriptors along with the source and destination data from the cache hierarchy between all test iterations. We ignore the descriptor allocation and preparation time for the reasons explained in \S~\ref{subsec:throughput_and_latency}. 
We leverage the Linux \texttt{Perf} tool~\cite{de2010new} to gather information on cycles and Intel's \texttt{pqos} library APIs~\cite{GitHubin30:online} to observe the LLC occupancy of individual cores, as well as isolating the evaluated cores to specific LLC ways for minimizing interference with other background tasks in the system.

\subsection{Throughput and Latency}\label{subsec:throughput_and_latency}

\begin{figure*}[!t]
    \begin{center}
    \includegraphics[width=0.8\linewidth]{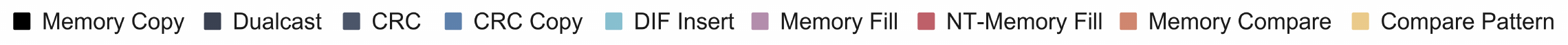}\end{center}
    \vspace{-26pt}
    \hfill 
    \vspace{-10pt}
    \\
    \centering
    \subfloat[Synchronous offloading\label{subfig:sync-no-batch-throughput}]{%
        \includegraphics[width=0.49\textwidth]{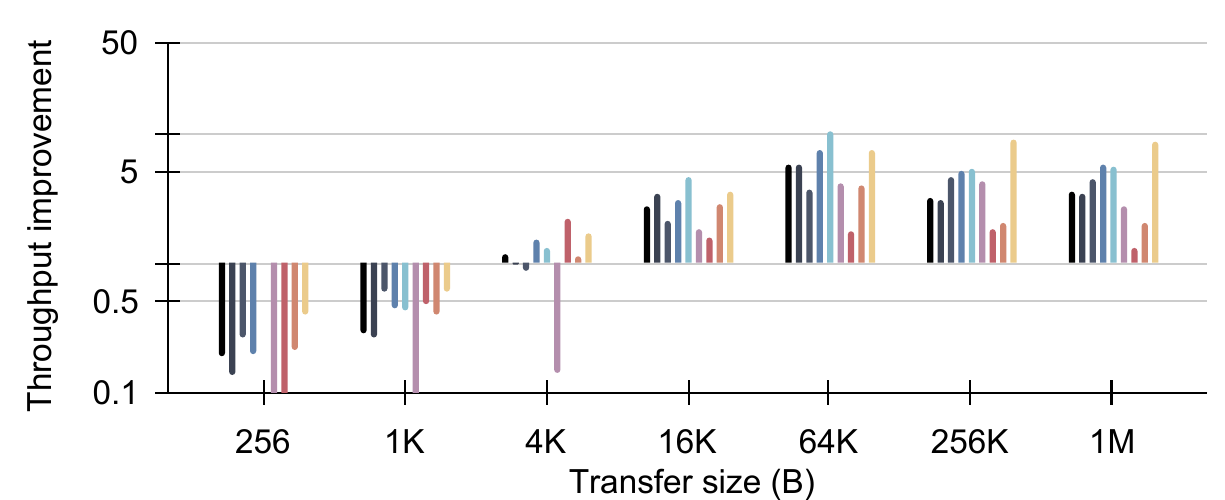}%
    }
    \hfill
    \subfloat[Asynchronous offloading\label{subfig:async-no-batch-throughput}]{%
        \includegraphics[width=0.49\textwidth]{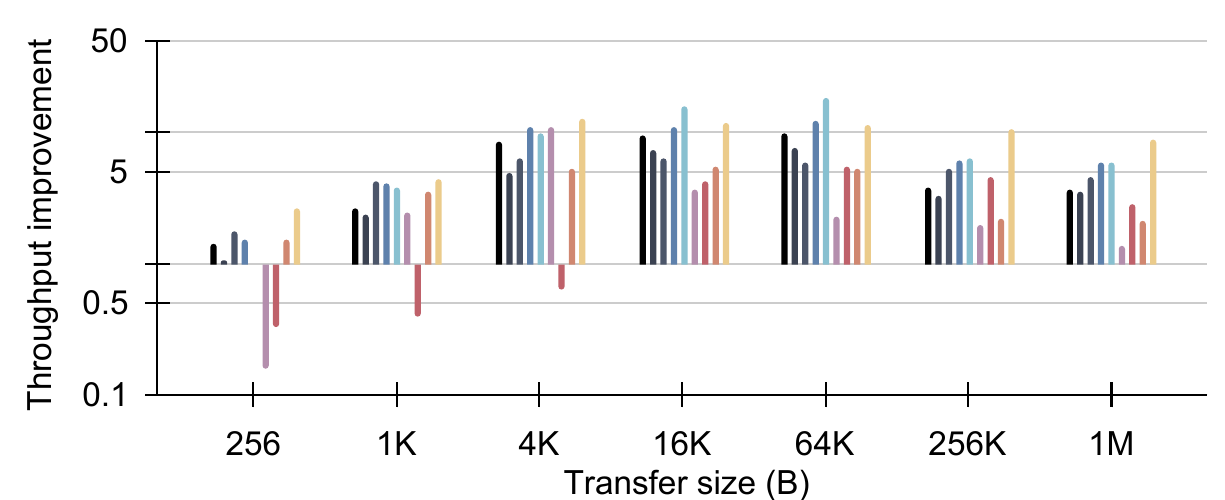}%
    }
    \vspace{-4pt}
    \caption{Throughput improvements of data streaming operations over their software counterparts
with varying transfer sizes (batch size: 1). \texttt{Memory Fill} and \texttt{NT-Memory Fill} refer to allocating and non-allocating writes (similar to regular \texttt{store} and \texttt{nt-store}), respectively.}
    \label{fig:no-batch-throughput}
    \vspace{-6pt}
\end{figure*}

\noindent \textbf{Comparison with CBDMA:}
Through conducted experiments, we found \accel to perform an average of 2.1$\times$ greater throughput than CBDMA found in Intel 3rd Gen Xeon scalable processors (ICX) over varying transfer sizes. This is mainly due to generational improvements such as the microarchitectural innovation of DMA engines and the scaling of the technology process, along with improvements in memory subsystem performance.

\noindent \textbf{Impact of transfer size:}
\fig~\ref{fig:no-batch-throughput} demonstrates the impact of data transfer size on the speedup of \accel over its software counterparts running on cores. In this experiment, we assume a pre-allocated memory space for work descriptors. In other words, the latency of \accel offloading includes only the time for submitting a descriptor and waiting for its completion, not for descriptor memory allocation. \accel shows favorable throughput improvements when offloading data above 4~KB in size when offloaded synchronously, where each descriptor is submitted and completed prior to sending another. These improvements above 4~KB are due to the reduced communication overheads for descriptor submission between the host CPU and \accel. This is also the case for the latency performance of \accel and CPU core, as demonstrated later.

The speedups presented so far are achieved when offloading a single descriptor at a time, but if applications are designed more carefully, we can batch multiple descriptors together as well as offload them asynchronously. \fig~\ref{subfig:async-no-batch-throughput} highlights the improvements when offloading work asynchronously, surpassing their software counterparts much earlier with transfer sizes around 256~bytes.

\begin{figure}[!t]
    \centerline{\includegraphics[width=0.9\columnwidth]{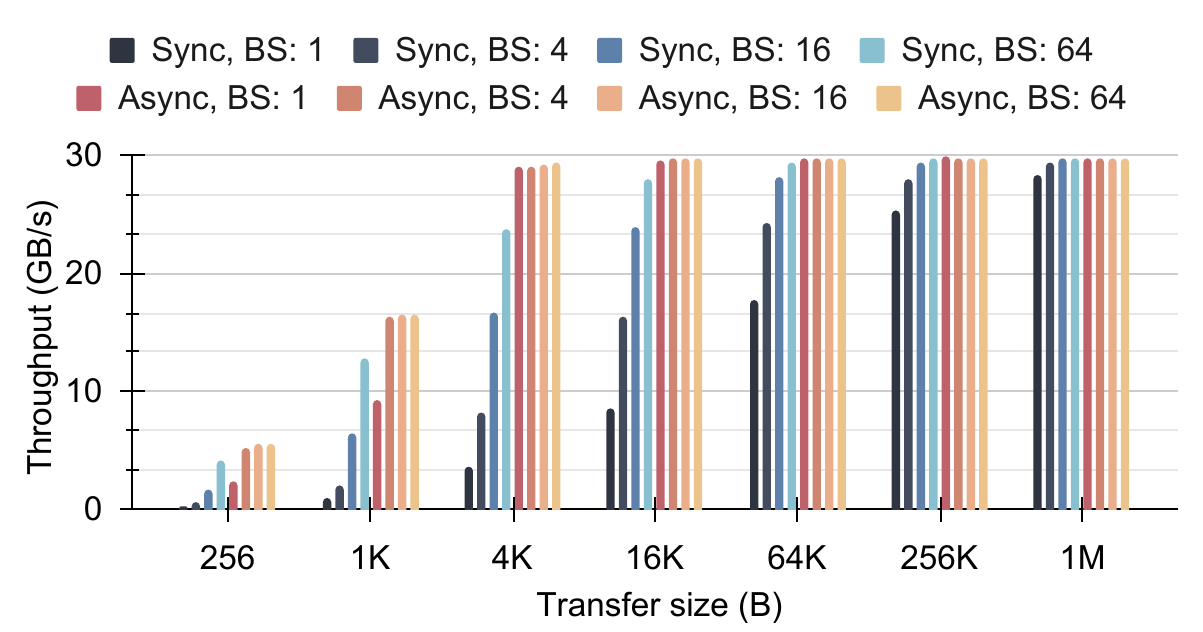}}
    \vspace{-4pt}
    \caption{Throughput of \accel's \texttt{Memory}~\texttt{Copy} operation on \texttt{Sync} or \texttt{Async} offloading with varying transfer sizes and batch sizes (\texttt{BS})}
    \label{fig:sync-async-ts-vs-bs}
    \vspace{-6pt}
\end{figure}

\noindent \textbf{Impact of batch processing:}
\accel allows descriptors to be batched to minimize the cost of offloading. 
\accel increases throughput non-linearly depending on the transfer size and level of synchronicity. For synchronous usage, the increasing batch size for small transfer sizes exponentially increases observed throughput, while more modest improvements are seen for transfer sizes above 256~KB as shown in \fig~\ref{fig:sync-async-ts-vs-bs}.

When varying both transfer and batch sizes, throughput is primarily affected by the asynchronous use of \accel. For streaming asynchronous work submission to a dedicated work queue (DWQ), a core can stream descriptors at a high enough rate that allows the device to achieve peak throughput even without batching (BS: 1). For asynchronous submission to a shared work queue (SWQ), a batch of size n is roughly equivalent to streaming submission by n CPU cores. This is because of the round-trip latency involved with the \texttt{ENQCMD} instruction used to submit to the SWQ.
After synchronously offloading 256~KB and asynchronously offloading 4~KB, with batch sizes of 64 and 4 respectively, the throughput saturates at 30~GB/s due to the inherent I/O fabric limit.

\begin{figure}[!t]
    \vspace{4pt}
    \centerline{\includegraphics[width=0.9\columnwidth]{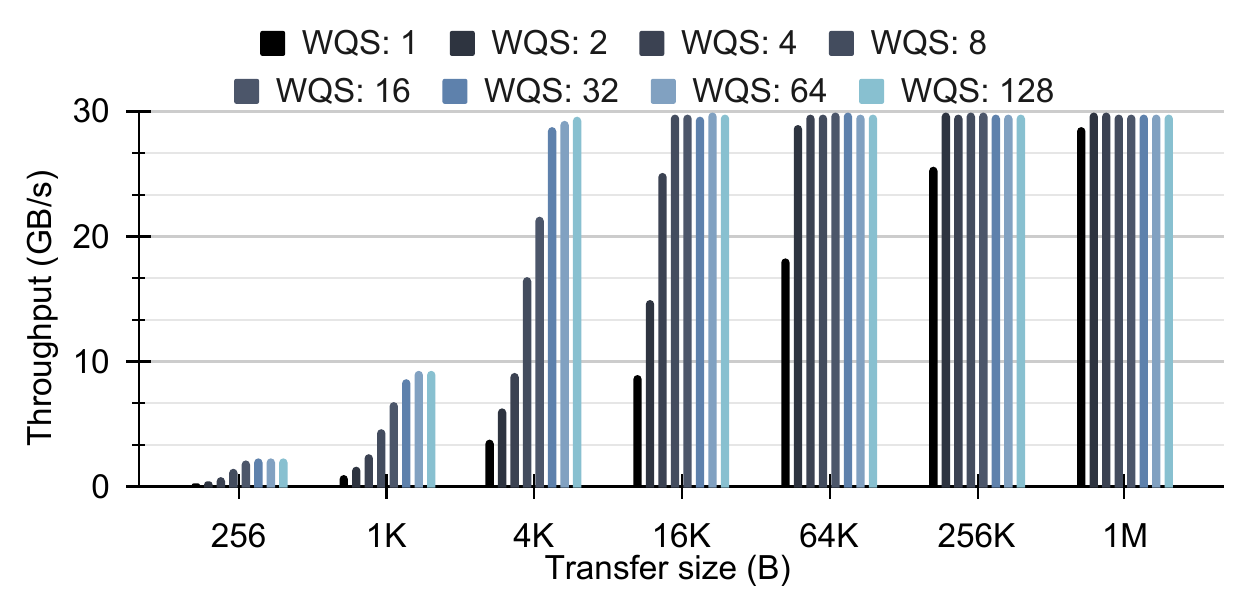}}
    \vspace{-6pt}
    \caption{Throughput of \accel's asynchronous \texttt{Memory}~\texttt{Copy} operation with different WQ sizes (\texttt{WQS})}
    \label{fig:work-queue-size}
    \vspace{-12pt}
\end{figure}

\begin{figure}[!t]
    \vspace{4pt}
    \centerline{\includegraphics[width=0.9\columnwidth]{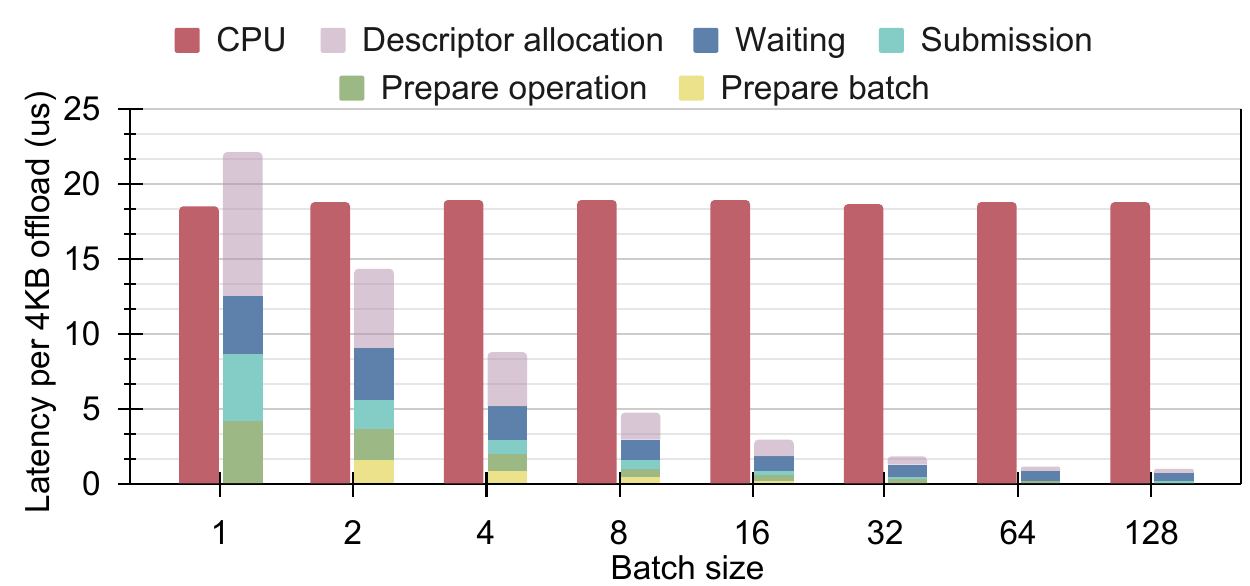}}
    \vspace{-4pt}
    \caption{Breakdown of \memcpy latency on CPU (left bars) and \texttt{Memory}~\texttt{Copy} operation latency on \accel (right stacked bars) with varying batch sizes (transfer size: 4~KB)}
    \label{fig:breakdown-4k-latency}
    \vspace{-6pt}
\end{figure}

\noindent \textbf{Impact of asynchronous use:}
As discussed earlier, asynchronous use of \accel can greatly improve throughput over the case where a single descriptor is offloaded at a time. This is because asynchronous submission of multiple descriptors allows overlapped execution and hiding of address translation and memory access latency through hardware pipelining. In \fig~\ref{fig:work-queue-size}, we observe that increasing the number of in-flight work descriptors (through increasing \texttt{WQ size}) improves the throughput of \texttt{Memory Copy} up to the saturation points by reducing the offloading costs. The same benefits are observed across all other operations. \fig~\ref{fig:work-queue-size} demonstrates the synergistic impact of batch processing and asynchronous offloading.

\begin{figure*}[!t]
    \centering
    \subfloat[Local (L) and remote (R) sockets\label{subfig:local-vs-remote}]{%
        \includegraphics[width=0.45\textwidth]{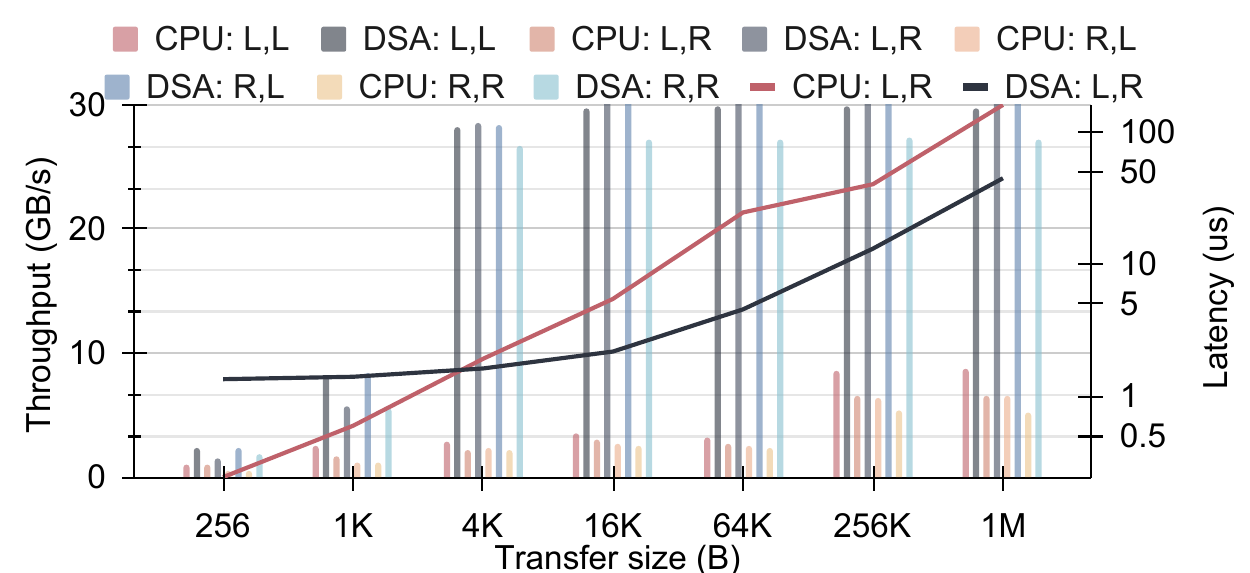}%
    }
    \hfill
    \subfloat[DRAM (D) and CXL (C)\label{subfig:dram-vs-cxl}]{%
        \includegraphics[width=0.45\textwidth]{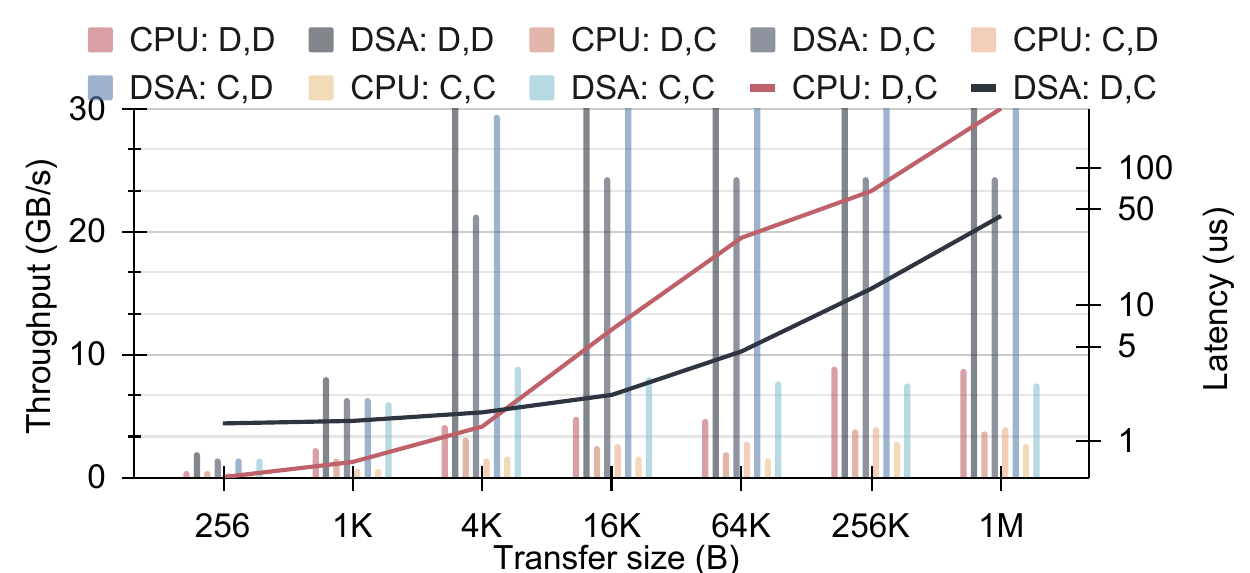}%
    }
    \vspace{-4pt}
    \caption{Throughput (bars) and latency (lines) impact when using different memory configurations (synchronous mode, batch size: 1). Each configuration is labeled as [<\highlight{Device}>: <\highlight{Source buffer}>,<\highlight{Destination buffer}>].}
    \label{fig:memory-configurations}
    \vspace{-6pt}
\end{figure*}

\noindent \textbf{Latency breakdown of \accel offload:}
Breaking down these latencies into separate steps in the offloading process makes it clear where the time is spent in the \accel workflow. We demonstrate the latency behaviors of both \accel (synchronous mode) and the CPU cores in \fig~\ref{fig:memory-configurations}, and further break the \accel latency down in \fig~\ref{fig:breakdown-4k-latency}. The time spent on \accel queueing or processing consumes the majority of the latency. The descriptor allocation time is where most time is spent, though in real-world use, these descriptors are often times pre-allocated in buffers ahead of use, amortizing the cost for this step, and can be ignored. 
Following allocation, the waiting and submitting time consume the next most significant times as this is where the computation and data reads are done. While waiting for the descriptor to be processed, the core becomes idle, allowing other applications to spend CPU cycles completing their tasks along with \accel operating in the background. Managing these tasks asynchronously with larger batches best utilizes the CPU cycles that would have otherwise been wasted on the common background tasks.  
Preparing the descriptor takes the least time, and as descriptor preparation would only assign the needed components (e.g., completion record, source and destination addresses, flags, etc.), these costs are evidently not as significant. In many applications, the values for flags and completion record addresses (if pre-allocated along with the descriptor) can be amortized from the process as well along with the descriptor allocation. Setting the status to zero can be done once the completion record is retrieved, allowing for only the source and destination addresses to be written to the descriptor regarding preparation. As this time involves two writes, it can considered low-cost and can be ignored.

\noindent \textbf{Impact of \accel across sockets:}
Memory access latency differs depending on whether data is located in memory local to the \accel socket or on the remote socket. As memory requests are sent and received by other sockets via Intel's Ultra Path Interconnect (UPI), the corresponding overhead increases remote memory latency. With pipelined execution and adequate buffering, \accel is generally able to effectively hide the additional latency to access remote socket memory. As a result, \accel is able to achieve full performance accessing remote memory similar to local accesses. The measured throughput and latency between the different memory configurations are shown in \fig~\ref{subfig:local-vs-remote} by assigning the source and destination buffers for the \texttt{Memory Copy} operation to the local or remote NUMA nodes.

The general trend indicates that latency breaks even with copying from the CPU between 4-10KB for all configurations and the example of using a local source buffer and remote destination buffer is added to \fig~\ref{subfig:local-vs-remote}. Also, allocating source/destination buffers in separate locations yield slight benefits as the system can take greater advantage of memory channel parallelism. 

\begin{figure}[!t]
    \centerline{\includegraphics[width=0.9\columnwidth]{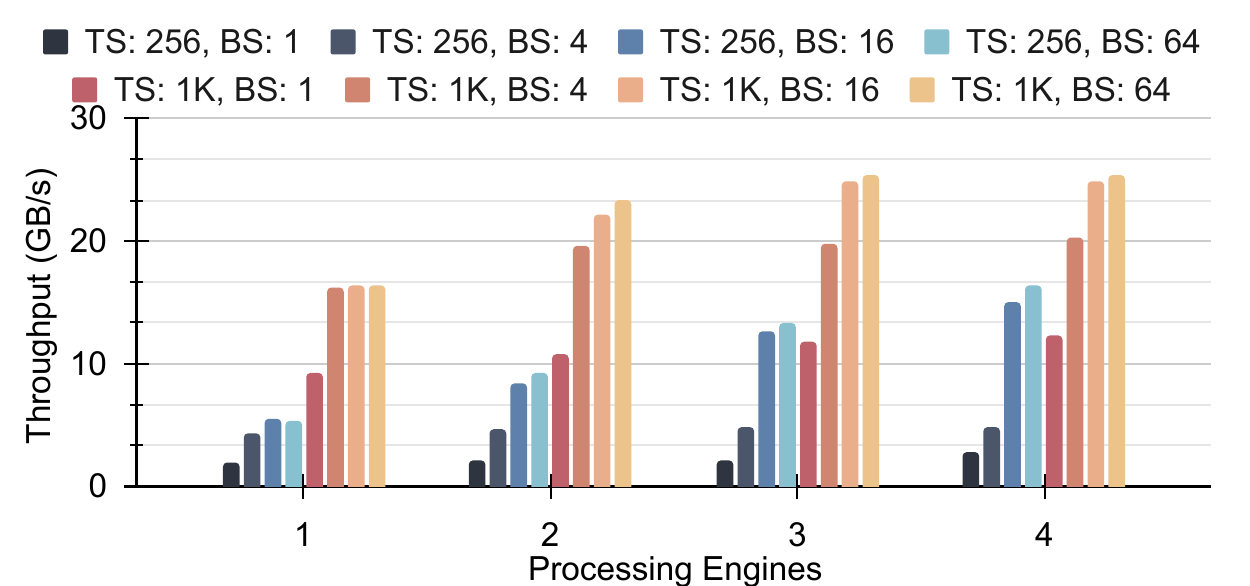}}
    \caption{Performance impact of the number of PEs on \texttt{Memory}~\texttt{Copy} with varying transfer sizes (\texttt{TS}) and batch sizes (\texttt{BS}) using one WQ}
    \label{fig:processing-engines}
    \vspace{-6pt}
\end{figure}
\noindent \textbf{Applying \accel to CXL-based memory:}
As an emerging interconnect technology, Compute eXpress Link (CXL)~\cite{cxl1,cxl2} has introduced a new type of memory device to the server's memory hierarchy. Typically, compared to the memory directly attached to the CPU's memory channels, CXL-based memory tends to have a larger capacity but lower latency/throughput performance. Thus, the cold data can be moved to it and brought back quickly when necessary. As Sapphire Rapids is the first Intel\textsuperscript{\textregistered} Xeon\textsuperscript{\textregistered} Scalable Processor that supports CXL (version 1.1), we measure and analyze \accel's performance on CXL-related data movement. We use an Intel\textsuperscript{\textregistered} Agilex-I Development Kit~\cite{intel-agi}, which is equipped with 16~GB DDR4 DRAM, as the CXL memory device. The CXL memory is exposed to the CPU/OS as a NUMA node with no CPU cores, and its usage is the same as regular NUMA memory management. 

Much like the cross-socket throughput, CXL-based memory accesses exhibit a similar, but far more exaggerated trend in \fig~\ref{subfig:dram-vs-cxl}. When using CXL memory for both source and destination buffers, throughput drops due to the higher access latency of CXL, but \accel still shows significantly higher throughput than its software counterpart. We also observe that the throughput is higher for configurations using CXL as the source and DRAM as the destination (compared to the opposite direction), due to the longer write latency of CXL-attached memory than its read latency.

\begin{figure}[!t]
    \centerline{\includegraphics[width=0.9\columnwidth]{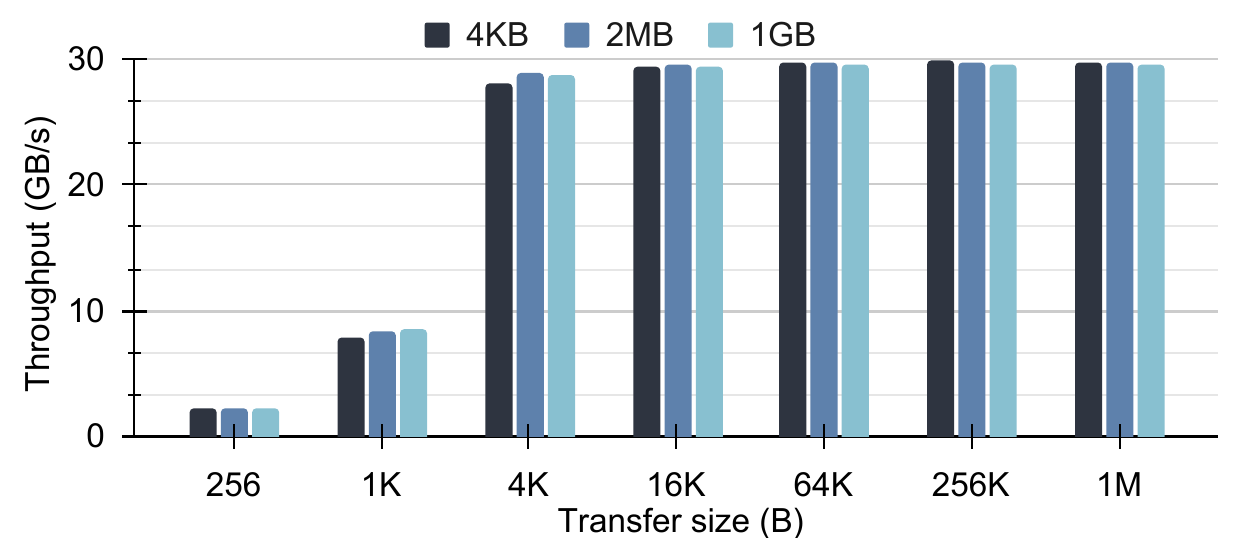}}
    \caption{Performance impact of using huge pages with varying transfer sizes.}
    \label{fig:huge-pages}
    \vspace{-6pt}
\end{figure}
\noindent \textbf{Impact of Huge Pages:}
Huge pages can improve the ability to cache data as TLBs cache larger regions of memory per single page table entry. This improvement comes at the cost of longer memory allocations and, when data is not spatially located, becomes a large overhead when page faults occur. The results from \fig~\ref{fig:huge-pages} show how throughput is nearly unaffected by the size of pages used.

\begin{figure}[!t]
    \centerline{\includegraphics[width=0.9\columnwidth]{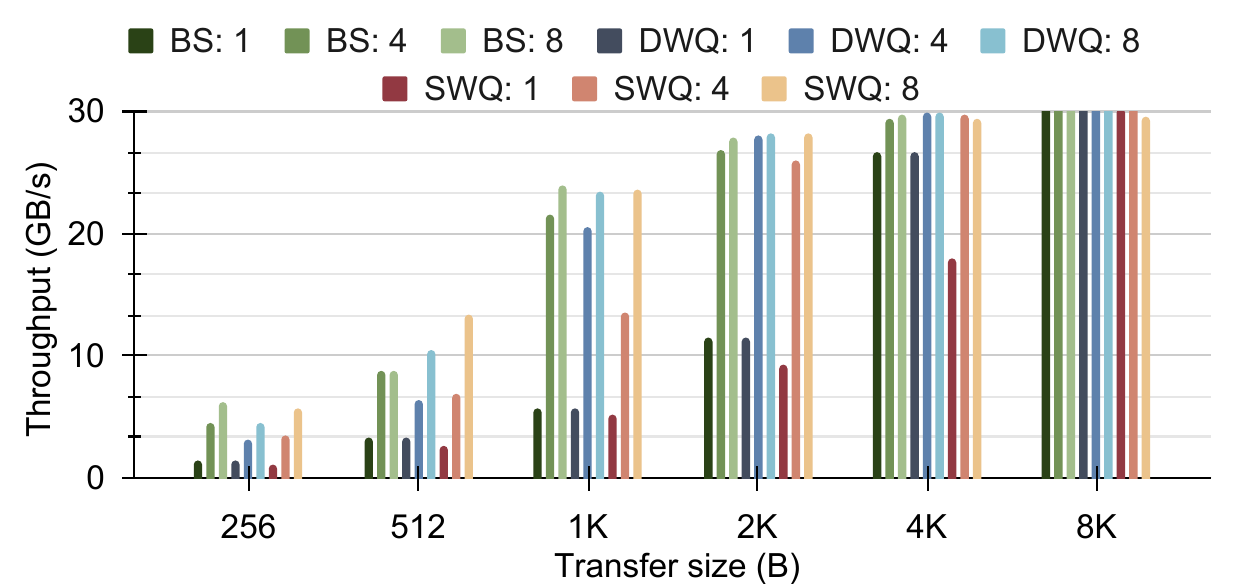}}
    \caption{Throughput impact of different WQ configurations: 1) one DWQ with batching (\texttt{BS: N}), 2) multiple DWQs with one thread and PE per queue (\texttt{DWQ: N}), and 3) one SWQ with one PE and multiple threads submitting (\texttt{SWQ: N}).}
    \label{fig:work-queue-configs}
    \vspace{-6pt}
\end{figure}

\begin{figure}[!t]
    \centerline{\includegraphics[width=0.9\columnwidth]{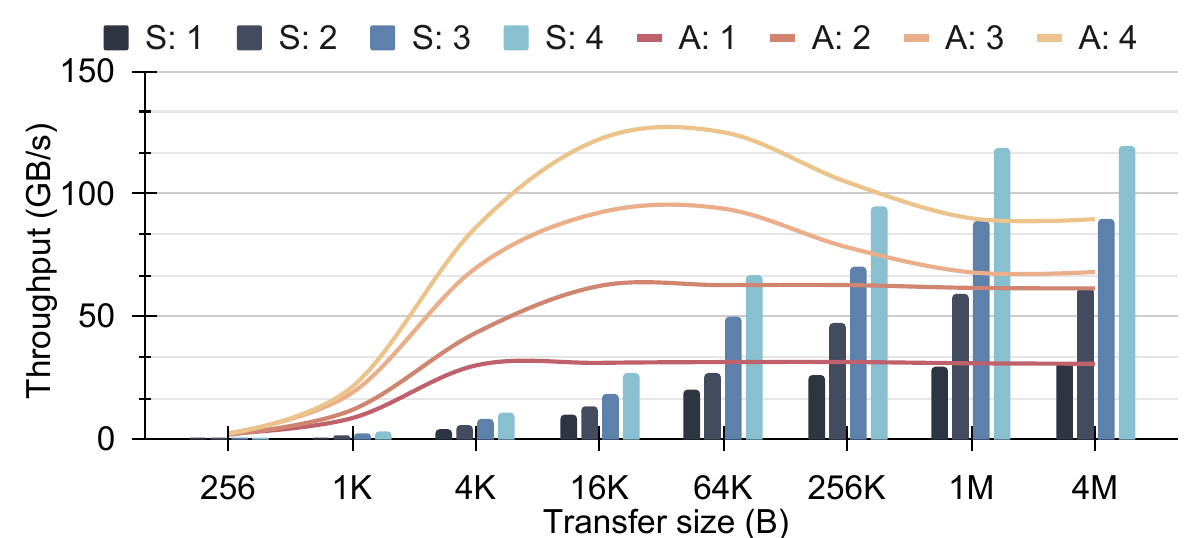}}
    \vspace{-4pt}
    \caption{Throughput using multiple DSA instances}
    \label{fig:multi-dsa-instances}
    \vspace{-6pt}
\end{figure}
\subsection{Flexible Group Configuration}\label{subsec:configurable_groups}
One of the unique characteristics of \accel is that users can flexibly configure the accelerator based on their needs. One primary configurable aspect is the number of PEs per group. When a group is equipped with a single PE, submitted descriptors are processed in a pipelined fashion and can reach throughput saturation on its own. However, if the PE in this scenario becomes stalled due to long-latency address translation or lengthy page fault handling, subsequent tasks submitted to this group are affected. On the other hand, when configured with more than one PE, a group can continue processing additional tasks as other free PEs in the same group are able to dispatch and process subsequent descriptors, providing higher throughput and QoS performance. This contributes to the improvements in \texttt{Memory}~\texttt{Copy} throughput with the increased number of PEs in \fig~\ref{fig:processing-engines}. For larger transfer sizes, configuring groups with more PEs show leveling improvements because a single PE is capable of hitting peak bandwidth in such cases.  

Groups may be configured to include multiple WQs, either shared or dedicated. \fig~\ref{fig:work-queue-configs} demonstrates the impact of offloading to different WQ configurations, including batching to a single DWQ, multiple DWQs, and one SWQ. Using either multiple DWQs or batching for a single queue results in nearly identical throughput, while SWQ observes lower throughput between 1-8~KB. When using many threads for offloading to a SWQ, the throughput matches that of the other WQ configurations.

Multiple \accel instances can be included on the same SoC, thus using multiple simultaneously can increase the maximum observed throughput. From \fig~\ref{fig:multi-dsa-instances}, throughput is shown to increase linearly with the number of \accel used. Beyond 64~KB, the maximum throughput begins to drop to 70~GB/s and 90~GB/s for 3 and 4 \accel instances respectively. This is because in such cases, the data write footprint exceeds the DDIO portion of the LLC (i.e., the leaky DMA problem~\cite{tootoonchian2018resq,yuan2021don}), hence, the throughput in such cases is limited by the the system memory bandwidth. In use cases demanding higher throughput with large transfer sizes, more LLC ways should be allocated for DDIO~\cite{yuan2021don} (also see \S~\ref{subsec:interation_with_cache_mem}).

\begin{figure}[!t]
    \centerline{\includegraphics[width=0.9\columnwidth]{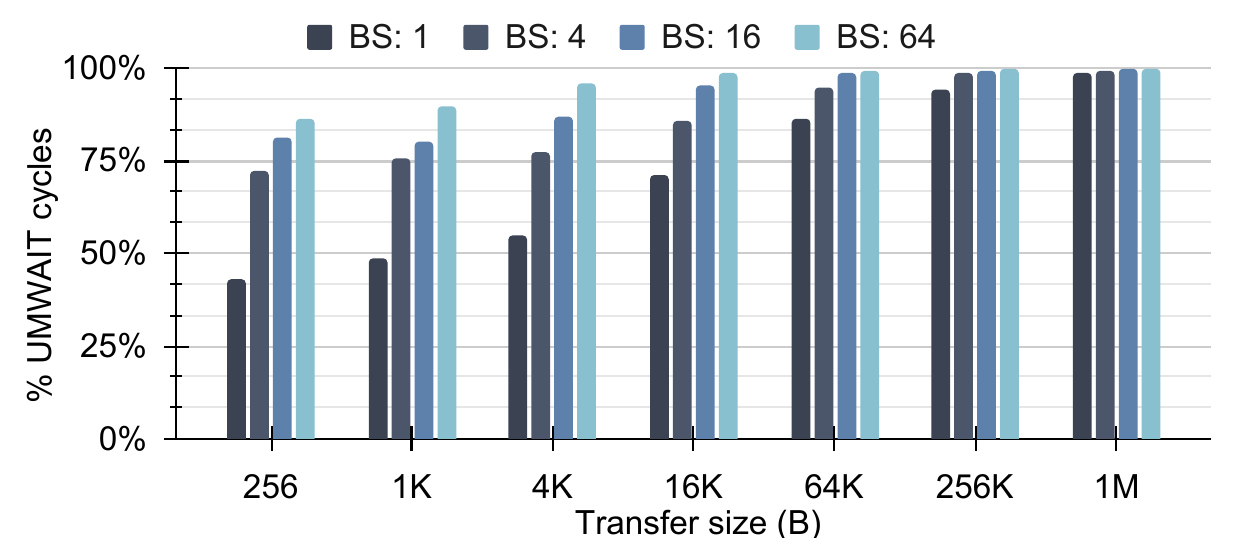}}
    \vspace{-6pt}
    \caption{CPU cycles spent on the \texttt{UMWAIT} intrinsic (and thus in a lower power state) when offloading \texttt{Memory}~\texttt{Copy} operations with varying transfer sizes and batch sizes (\texttt{BS}) }
    \label{fig:umwait-cycles}
    \vspace{-6pt}
\end{figure}
\subsection{CPU Cycle Reduction}\label{subsec:cpu-cycle-reduction}
Offloading work to \accel allows the core running the process to use an optimized sleep state via \texttt{UMWAIT} to save dynamic energy consumption (\S~\ref{subsec:dsa_sw_arch}). The user may also choose to use interrupts instead of \texttt{UMWAIT} to notify the offloading core when the descriptor is finished processing. During this processing time, the core can switch to work on other tasks as it waits for the operation to complete. \fig~\ref{fig:umwait-cycles} shows the percentage of CPU cycles spent on the \texttt{UMWAIT} intrinsic while offloading \texttt{Memory}~\texttt{Copy} descriptors. With the transfer sizes of 4~KB or higher, the majority of cycles are spent on \texttt{UMWAIT} intrinsic. Furthermore, when batched, most cycles are spent on \texttt{UMWAIT} across all transfer sizes, which indicates that the host CPU can leverage such saved cycles for performing other tasks.

\begin{figure*}[!t]
    \includegraphics[width=.95\linewidth]{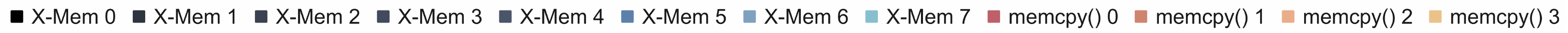}
    \vspace{-8pt}
    \hfill \\
    \centering
    \subfloat[None\label{subfig:xmem-small-none}]{%
        \includegraphics[width=0.32\textwidth]{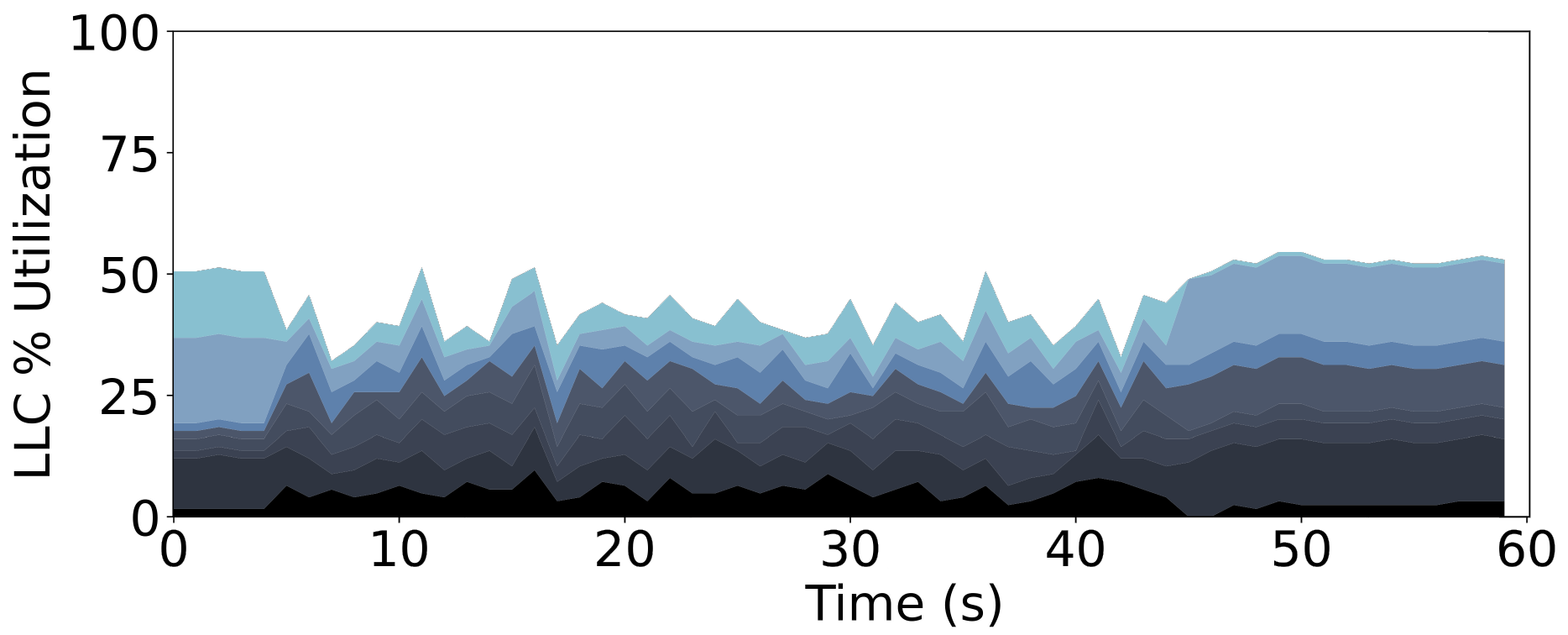}%
    }
    \hfill
    \subfloat[Software\label{subfig:xmem-small-cpu}]{%
        \includegraphics[width=0.32\textwidth]{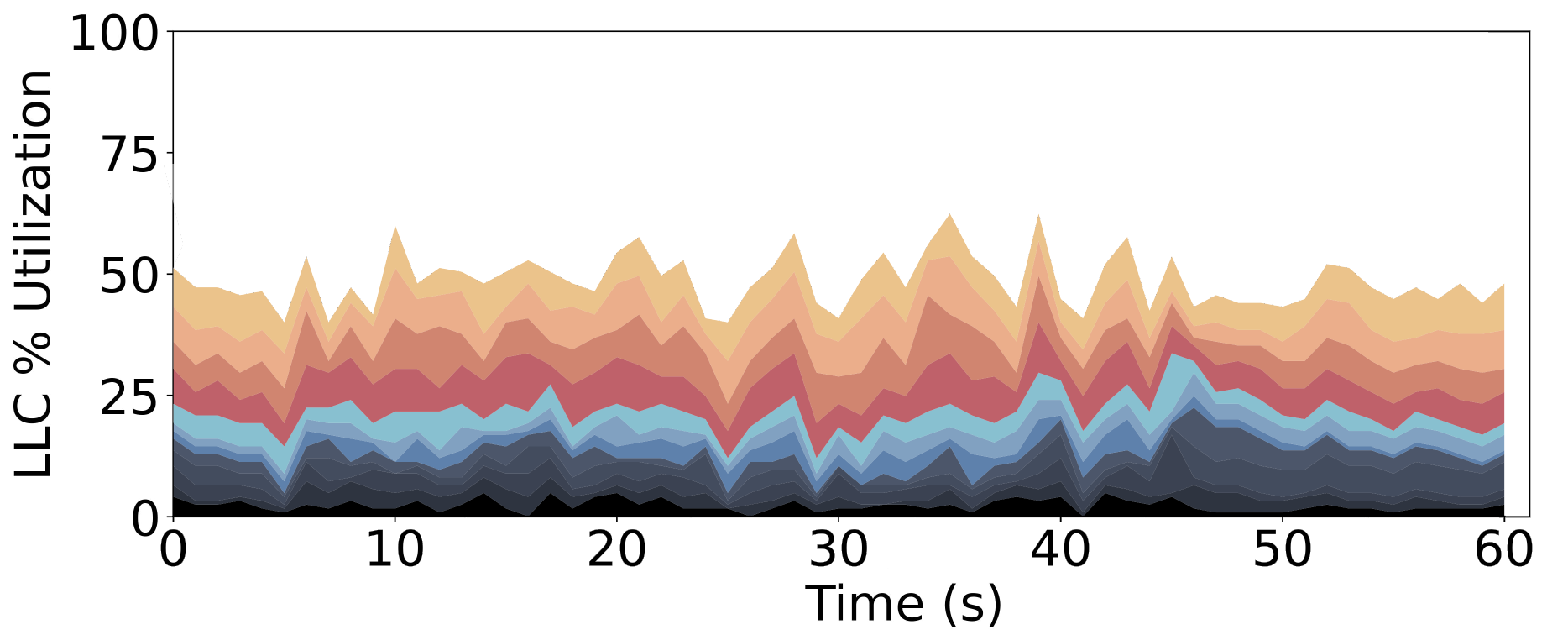}%
    }
    \hfill
    \subfloat[\accel offload\label{subfig:xmem-small-dsa}]{%
        \includegraphics[width=0.32\textwidth]{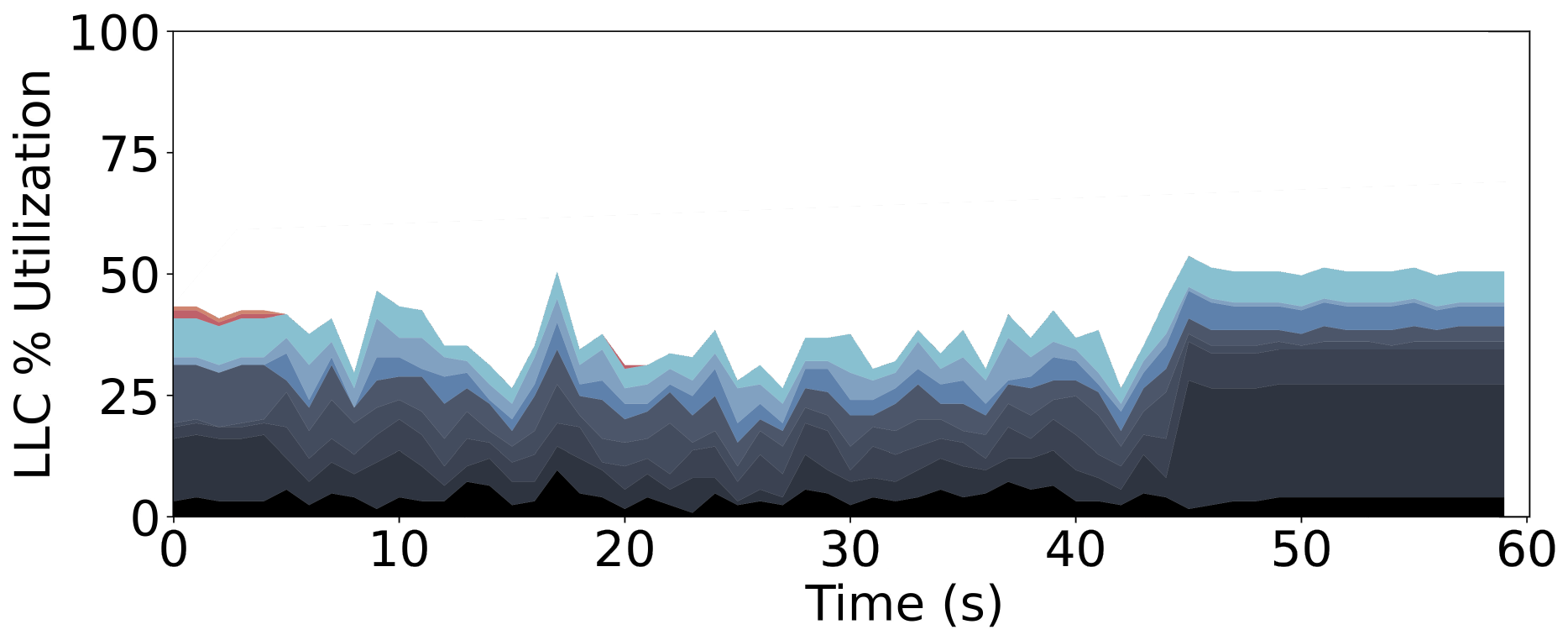}%
    }
    \caption{LLC occupancy of each individual core running either an \texttt{X-Mem} (black to light blue) or \memcpy (red to orange) process on three co-running scenarios in \fig~\ref{fig:xmem-read-latency} (focusing on 4~MB working set)}
    \label{fig:xmem-small}
    \vspace{-8pt}
\end{figure*}

\begin{figure}[!t]
    \centerline{\includegraphics[width=0.9\columnwidth]{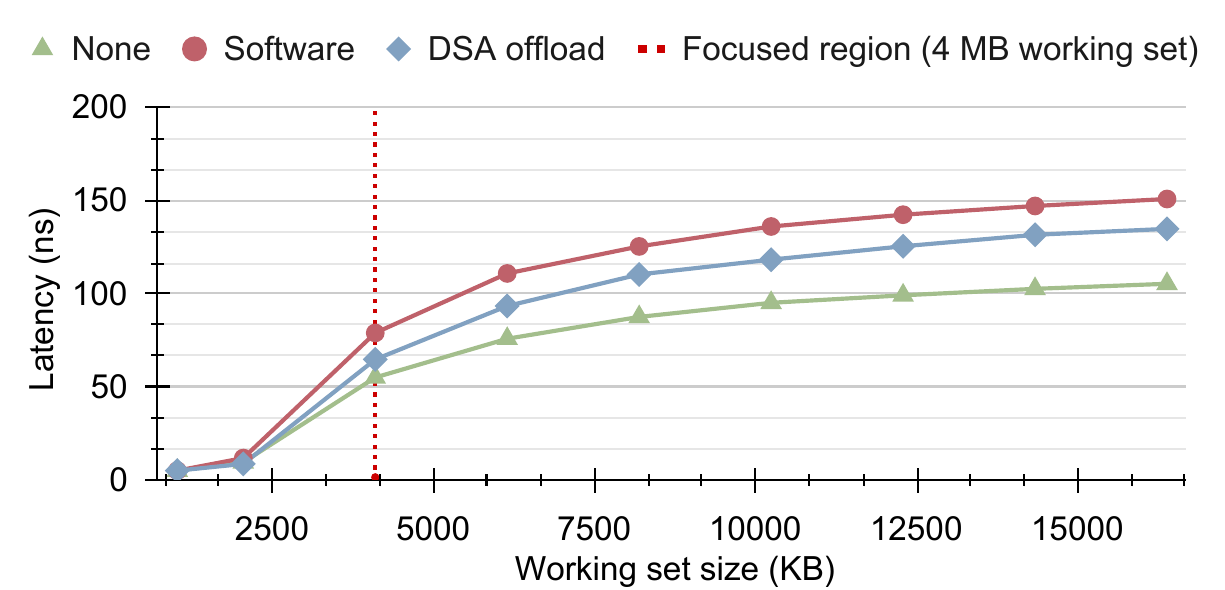}}
    \caption{Latency test results of running eight \texttt{X-Mem}~\cite{xmem} instances with varying working set sizes. Three different co-running application scenarios are presented: (1) \textbf{None}: no co-located processes, (2) \textbf{Software}: four \memcpy processes are running on separate cores (transfer size: 4~KB), (3) \textbf{\accel offload}: \texttt{Memory}~\texttt{Copy} operations are offloaded to four \accel groups (transfer size: 4~KB, batch size: 128).}
    \vspace{-6pt}
    \label{fig:xmem-read-latency}
\end{figure}

\subsection{Cache Pollution}\label{subsec:cache_pollution}
One of the key benefits of offloading tasks to \accel is to prevent the LLC space from being polluted by the streaming data having low locality. This allows other co-located, latency-sensitive applications to meet their desired timing. To demonstrate this, we run a memory-intensive microbenchmark, \texttt{X-Mem}~\cite{xmem}, and evaluate performance in terms of average read/write latency while running background tasks that perform memory copy operations. \texttt{X-Mem} is a micro-benchmark that allows users to specify the working-set size (i.e., memory footprint) and estimate the average throughput and access latency with specific access patterns (e.g., sequential and random, reads and writes, etc.) on the memory region. 

In \fig~\ref{fig:xmem-read-latency}, we run eight \texttt{X-Mem} instances with no background processes (\texttt{None}), four \texttt{memcpy()} processes running on four separate cores (\texttt{Software}), and four \texttt{Memory}~\texttt{Copy} tasks offloaded to \accel (\texttt{Intel}~\texttt{DSA}~\texttt{offload}). Even though each process runs on a dedicated core, \texttt{X-Mem} sees an increased memory access latency with co-located \texttt{Software} (e.g., 43\% longer than \texttt{None} with 4~MB working set size). This is mainly due to the cache pollution caused by streaming data copied by \texttt{Software}, which is significantly mitigated by offloading \texttt{Memory}~\texttt{Copy} tasks to \accel. With \accel, reads do not allocate into cache, and writes are limited to DDIO portion of the LLC~\cite{yuan2021don}, thus significantly minimizing cache pollution effects. \fig~\ref{fig:xmem-small} further investigates this by keeping track of the LLC occupied by individual cores. In this experiment, \texttt{X-Mem} instances run from 5s to 45s, while the background memory copy operations run from 0s to 60s. In \fig~\ref{subfig:xmem-small-cpu}, \memcpy processes dominate the LLC occupation. In contrast, there is almost no LLC occupation when using \accel as the background tasks operate only on data in memory without caching.

\section{\accel Software Enablement and Ecosystem}
\label{sec:use-cases-and-sw-support}
The available operations supported by \accel and its substantial throughput improvements and speedups, demonstrated in \S~\ref{sec:result_and_analysis}, bring many possible use cases. In addition to the basic IDXD driver and \texttt{libaccel-config} API library, software enablement efforts for \accel have been made to build the \accel ecosystem. Through active community development, software stacks for data-intensive scenarios (e.g., networking, storage, caching) have had proper support for \accel. In this section, we describe such efforts in multiple critical areas, as well as a couple of common software libraries for \accel enablement.
See the appendix for three specific examples -- SPDK NVMe/TCP Target, CacheLib-based Cloud Data Caching Service, and \texttt{libfabric} in HPC/ML.

\noindent \textbf{Network acceleration:}
The benefits I/OAT brought to network acceleration are amplified through \accel's greater throughput and more supported operations to assist networking. Offloading packet copying in virtualized networking stacks, such as virtIO, is one example (see \S~\ref{subsec:casestudy} for detailed case study). In conjunction with packet copying for intra-VM/container communication~\cite{10.1145/3555050.3569118}, accelerating other networking stacks like VPP~\cite{vpp}, and networked applications like video transport~\cite{vtp},  through \accel brings increased throughput and computation per core, while alleviating processing strain on the overall system.

\noindent \textbf{Storage Acceleration:}
\accel's characteristic is also a natural fit for storage stacks, where large bulks of data need to be moved and checked. 
Along with SPDK's software stack for \accel enablement (in appendix), technologies like \texttt{io\_uring}~\cite{iouring} and Distributed Asynchronous Object Storage (DAOS)~\cite{daos:online} offer high performance by removing common block-based I/O bottlenecks through using new storage interface/semantics. HPC workloads using DAOS have the potential to leverage \accel's memory mover capabilities to improve memory throughput, reduce core strain on simple data tasks, and coalesce small I/O requests in the background~\cite{daos-dsa-accel:online}.

\noindent \textbf{Datacenter tax reduction:}
As identified by Google and other cloud giants~\cite{Kanev:2015:PWC:2749469.2750392}, lots of CPU cycles in the cloud datacenter are not spent on the applications it is carrying -- instead, the cloud infrastructures themselves take up these cycles. Such datacenter tax includes but is not limited to device and memory management, virtualization, RPC stack, etc.
\accel's rich functionalities have been able to reduce datacenter taxes by offloading routines in memory compaction,  VM/container boot-up and migration, and more key scenarios.
    
\noindent \textbf{HPC/ML acceleration in datacenters:}
In addition to the system infrastructural use cases, \accel can also be applied in a wide range of applications. One example is machine learning (ML) processing frameworks. On the one hand, as an iterative process, machine learning processing requires clearing certain memory regions (e.g., gradient vectors) before it can proceed to the next iteration; on the other hand, in the scenarios of distributed ML, a huge amount of tensors and vectors need to be copied and transferred across nodes in a cluster via communication libraries like MPI and \texttt{libfabric}~\cite{7312664} (in appendix). As ML model sizes increase rapidly over the years, such operations can create a significant burden to the CPU. With \accel's assistance, such memory-zeroing and memory-copy routines have been offloaded and the CPU cycles can be saved for more critical tasks. Note that similar problems also occur in certain HPC workloads, where \accel can be leveraged. 

\noindent \textbf{Data movement offloading for disaggregated memory systems with tiered CXL:}
CXL-based memory expansion has emerged as the industry's de-facto memory expansion and disaggregation solution. Being more flexible with higher memory capacity, CXL-based memory may have longer access latency compared to the regular DRAM-based main memory. This makes the \texttt{LD/ST} semantics inefficient for the core to move chunks of data across different memory mediums, which is common in tiered memory systems. By offloading such memory movement and manipulation operations to \accel, we can enjoy the benefits of CXL memory, while not hurting the core's performance by saturating the load-store queue.

\noindent \textbf{Software libraries for \accel:}
In general, \accel enablement requires case-by-case software source code changes and recompiling for the applications. 
To increase the ease-of-use of \accel, we developed Intel Data Mover Library (DML)~\cite{dml} and \accel Transparent Offload library (DTO) for different levels of \accel usage. DML provides a set of high-level C/C++ APIs for data movement and transformation, calling the underlying \accel unit when available. DML supports advanced capabilities (all \accel hardware operations, asynchronous offload, load balancing, etc.), and applications can explicitly call DML APIs to take full advantage of DSA offload operations. DTO is a less intrusive library that allows the application to leverage DSA transparently without source code modification. For applications that wish to use the DTO library, a user can either dynamically link the library by using the linker options \texttt{-ldto} and \texttt{-laccel-config}, or preload the library via \texttt{LD\_PRELOAD} without having to recompile their application. When common system API calls like \memmove, \memcpy, \memset, or \memcmp are used, DTO functions by intercepting and replacing them with corresponding synchronous \accel operations. 

Meanwhile, \accel performance telemetry functionalities are also provided by the PCM library~\cite{pcm}. By reading the hardware performance counters, PCM is able to observe the inbound/outbound traffic and request count on each \accel instance.

\section{Make the Most out of \accel}\label{sec:discussion}
Having a comprehensive performance analysis and software enablement experience on \accel, we now summarize our \textbf{\underline{G}}uidelines on how to take advantage of \accel from three aspects: maximizing throughput, interaction with cache/memory hierarchy, and configuration of DSA hardware resources.
Then, applying these guidelines, we use DPDK-based virtIO as an example to demonstrate how \accel can be adopted optimally with notable performance benefits --- higher throughput and lower tail latency.

\begin{figure}[!t]
    \centerline{\includegraphics[width=0.9\columnwidth]{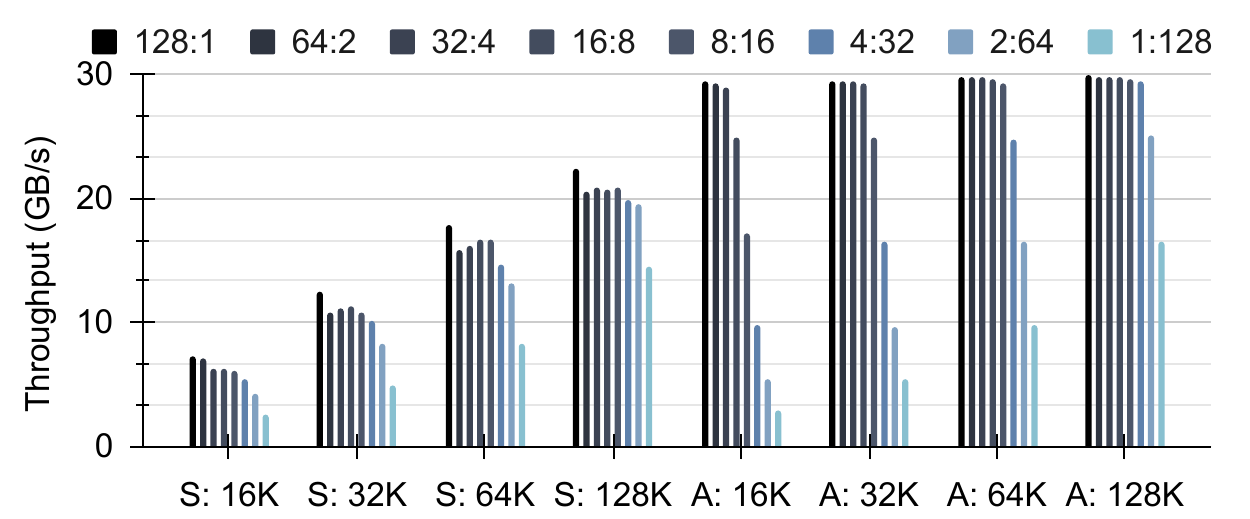}}
    \caption{Throughput of data streaming operations with varying transfer sizes per descriptor (\texttt{TS}) and batch sizes (\texttt{BS}). Legend format: <\texttt{TS}:\texttt{BS}>. X-axis format: <\texttt{Sync/Async}:\texttt{total transfer size}>.}
    \label{fig:equal-offload}
    \vspace{-6pt}
\end{figure}

\subsection{Optimizing Software Programming Models}
As an on-chip accelerator for streaming data, \accel requires the software programmer to tune the programming models for optimal performance.

\niparagraph{G1: Keep a balanced batch size and transfer size.} 
Offloading work of a certain size to \accel can be done by either using one descriptor for the full memory region or batching multiple smaller descriptors for the same aggregate size. \fig~\ref{fig:equal-offload} shows the relationship between equivalent offloading sizes and changing the ratio of transfer size and batch size. While not as impacted as changing other aspects of \accel, the general trend indicates a decrease in throughput when using larger batches for the same overall offloaded work. As individual descriptors must be processed internally in \accel and read the corresponding data from memory, the additional overhead for managing the increased number of descriptors may reduce the effective throughput achieved from these operations. If the desired data for offloading is contiguous, coalescing into a larger single descriptor of the equivalent size may improve both throughput and latency.

When offloading synchronously, a weak pattern emerges showing an optimum point in throughput between maximizing batches and maximizing transfer size. From \fig~\ref{fig:equal-offload}, modestly batching the work (4$\sim$8 descriptors) yields the best results. This is  due to balancing the latency from fetching sequential regions of memory and processing batched descriptors.

\begin{figure}[!t]
    \centerline{\includegraphics[width=0.9\columnwidth]{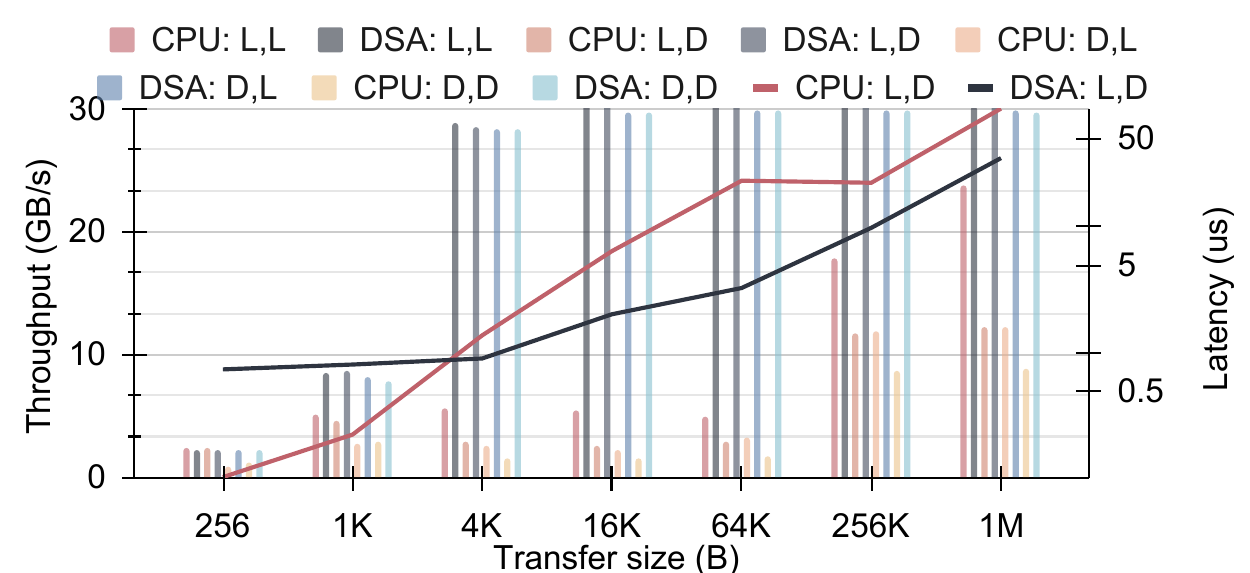}}
    \caption{Throughput (bars) and latency (lines) impact when offloading data from/to either LLC (L) or local DRAM (D) (batch size: 1). Labeling scheme is the same as \fig~\ref{fig:memory-configurations}.}
    \label{fig:dram-vs-llc}
    \vspace{-6pt}
\end{figure}

\niparagraph{G2: Use \accel asynchronously when possible.}
From \fig~\ref{fig:no-batch-throughput} and \fig~\ref{fig:dram-vs-llc}, offloading operations to \accel in an asynchronous manner provides optimal efficiency and performance for both the CPU core and \accel hardware. This can either use DML for quick implementation, or a reworking of an application. When limited in asynchronous potential, transfer sizes below 4~KB should be used on the CPU core if cache pollution is acceptable.

\subsection{Interaction with Cache/Memory Hierarchy}
\label{subsec:interation_with_cache_mem}
As an on-chip accelerator, \accel brings more options  to interact with the cache and (heterogeneous) memory hierarchy. 

\niparagraph{G3: Control the data destination wisely.} Unlike the completion record that is always directed to the LLC, data written to the destination address of a descriptor can be steered either to the LLC or to the main memory. \accel facilitates this \textit{cache control} feature by allowing users to provide a hint (i.e., setting the cache control flag of a work descriptor) that notes the preferred destination. If the flag is set to 0, the data is written to the memory while invalidating the corresponding cache lines in the LLC, if any. If the flag is set to 1, the data is directly written to the LLC by allocating the corresponding cache lines. The underlying principle of this technique is identical to that of Intel's Data Direct I/O (DDIO) technology, a direct cache access (DCA) scheme leveraging the LLC as the intermediate buffer between the processor and I/O devices~\cite{IntelDa26:online,huggahalli2005direct,alian2020data,alian2022idio}.

Similar to what was shown in \fig~\ref{fig:xmem-small}, cache pollution causes negative effects on co-running processes that share limited hardware resources. In instances of many datacenter workloads, the latency gained from antagonistic background cache evictions may prove to undermine the competitive service level agreements (SLAs) set for the primary applications. On the other hand, writing data which is either critical to performance or used by the core in the near future directly to the cache will provide an access latency and throughput advantage for the application, as shown in \fig~\ref{fig:multi-dsa-instances}. The programmer should wisely choose the data destination based on the application behavior.

\niparagraph{G4: \accel as a good candidate of moving data across a heterogeneous memory system.}
Moving data from/to different memory mediums, such as NUMA remote memory, persistent memory, and CXL-based memory, is common in modern tiered memory systems~\cite{258860,10.1145/3419111.3421294,10.1145/3477132.3483550,pmem-patch}. 
As noted in \S~\ref{sec:result_and_analysis}, using \accel to move or transform data from local or remote memory can affect throughput. Using memory with different characteristics like DRAM and CXL-attached memory displays similar results, but with more drastic effects. The ideal case in most scenarios favors having data local and on faster memory, but when that option is not feasible due to available memory capacity or the location of processes accessing the data, two patterns emerge. When using memory of similar bandwidth and performance, allocating source and destination addresses to separate locations yields marginally greater throughput as seen in \fig~\ref{subfig:local-vs-remote}. In cases where available memory types differ in their characteristics, the memory type with faster write latency exhibits better performance when used as \accel destination. CXL memory has lower latency for reads compared to writes and thus experiences greater effect when the written destination buffer is located on faster local DRAM. \fig~\ref{fig:dram-vs-llc} shows the difference in throughput and latency when data is either located within the offloading CPU's LLC or in local DRAM. It appears beneficial to offload using larger transfer sizes to \accel (\ie 4~KB when synchronous, and 128~B when asynchronous assuming no batching). Conversely, smaller transfer sizes may be better processed on the core if cache pollution would not impact other co-running applications.

\subsection{Configuration of \accel Hardware Resources}
\accel offers a high degree of configuration flexibility through the available hardware resources. Taking advantage of such configurations can yield optimal \accel hardware utilization, and thus better performance.

\niparagraph{G5: Leverage PE-level parallelism.}
As noted in \fig~\ref{fig:processing-engines}, the number of PEs used in a group can impact the maximum observed throughput. Increasing the number of PEs per group improves throughput. Users should be aware of the common transfer size of offloaded tasks to these groups as smaller transfer sizes yield greater performance scaling.

\niparagraph{G6: Optimize WQ configuration.}
Using batching or DWQs provide greater benefits compared to a SWQ. Unless the SWQ is in use by many other threads, greater utilization and throughput can be achieved through re-configuring to using more DWQs. SWQs may perform worse when few threads are used but can outperform all configurations when using more threads than the total number of WQs, as it offloads concurrency management to hardware. Additionally, WQs can be configured to either shared or dedicated for providing performance isolation between WQs within a group.

As \accel has limited WQ entries, assigning 32 entries for a single WQ can provide almost the maximum throughput possible.

\begin{figure}[!t]
    \centering
    \subfloat[Design overview\label{subfig:dpdk-based-vhost-dsa-flow-and-diagram}]{%
        \includegraphics[width=0.45\textwidth]{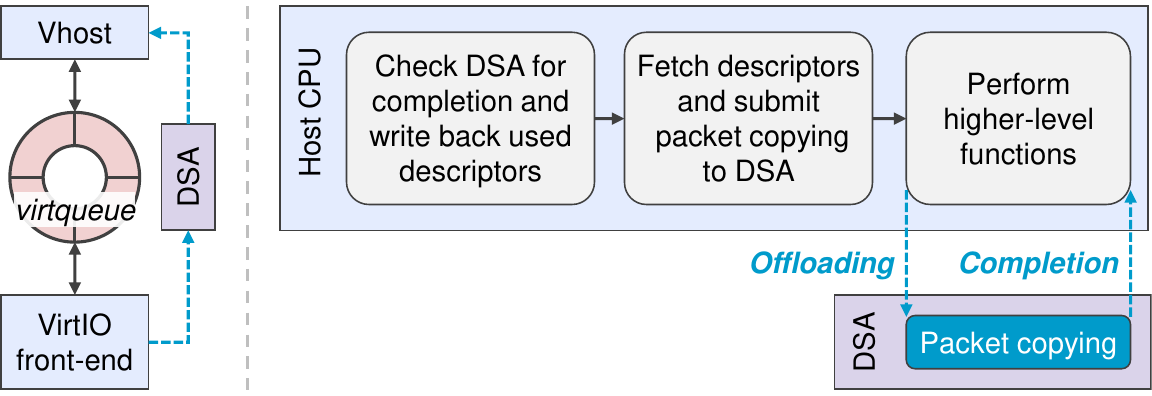}%
    }
    \hfill
    \vspace{-12pt}
    \subfloat[Throughput of packet forwarding\label{subfig:dpdk-based-vhost-dsa-accel}]{%
        \includegraphics[width=0.44\textwidth]{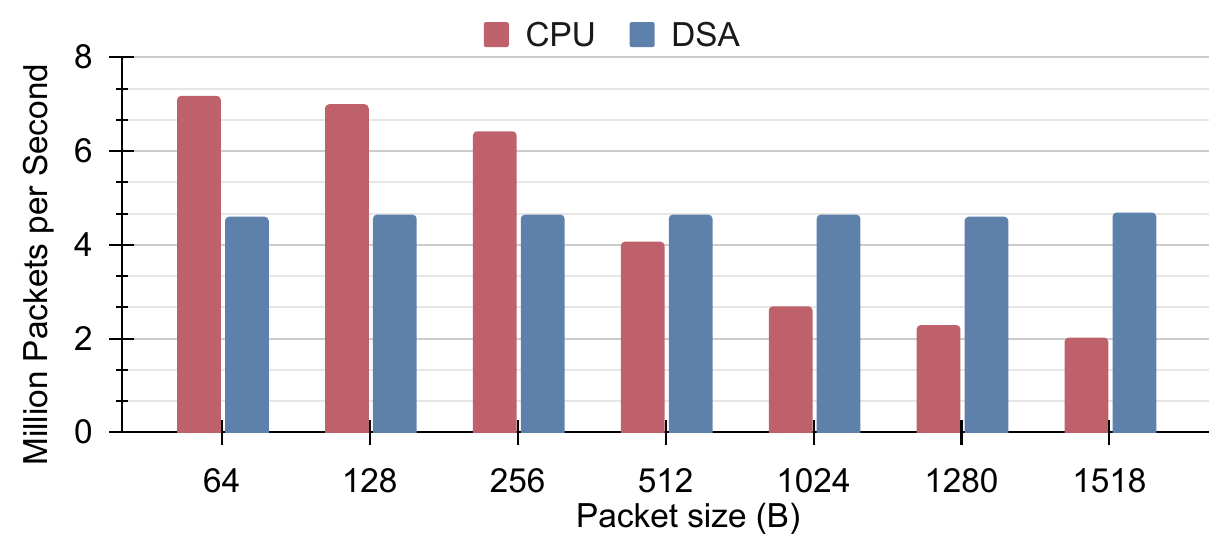}%
    }
    \vspace{-4pt}
    \caption{Accelerating packet copy operations in Intel DPDK Vhost library via \accel} 
    \label{fig:dpdk-based-vhost-with-dsa}
    \vspace{-9pt}
\end{figure}

\subsection{Case Study: DPDK-Based VirtIO}
\label{subsec:casestudy}
DPDK Vhost is a well-optimized and efficient VirtIO backend commonly used in many software switches, such as Open Virtual Switch (OvS), for managing data between a host backend and a VM frontend. Despite the improvements Vhost provides, large packet sizes reduce performance as packet copying becomes the dominant subprocess where CPU cycles are spent. In fact, according to our performance analysis, packet copying accounts for 30\% of CPU cycles when packet sizes are at 512 bytes and nearly 50+\% for any packet sizes larger than 1024 bytes. Consequently, the packet forwarding rate for such large packets can drop significantly as high as 38\% as packet size goes from 256B to 1KB. This makes DPDK Vhost a great candidate for packet copy offload to \accel. We describe the journey of applying \accel in DPDK Vhost software stack as an example of efficient and optimized \accel software enablement.

To better understand this offloading opportunity, we take a closer look at DPDK Vhost's mechanism of packet forwarding, where packet copying is intensively involved. In general, DPDK Vhost leverages the descriptor queue (\texttt{virtqueue}) data structure to manage packet transfer. 
An enqueue operation of virtqueue includes the following three steps.
(1) From the designated virtqueue, we fetch the available descriptors that describe buffers in VM memory. Then, from those descriptors, the addresses of empty buffers, which will store the future packets, are read for later packet operations. 
(2) We copy the packets from the host to the obtained buffers.
(3) We write back the “used” descriptors to the virtqueue and send a notification to the VM.
Similarly, a dequeue operation includes these three steps, but in a reverse order. 
Among these three steps, two memory copies are involved, one for used descriptor write-back and the other for packet copying. 

Obviously, the descriptor write-back is not worth offloading to \accel due to its small size ($\sim$10 bytes). Also, we realize that merely replacing \memcpy with the corresponding \accel calling functions may not yield tangible performance benefits due to \accel's long invoking latency. Hence, following the insights and guidance in the previous sections, we apply several software optimizations around \accel to amortize its overhead and amplify the offloading benefits.

As indicated by \textbf{G2}, we design an asynchronous programming model in virtqueue's enqueue and dequeue operations, as depicted in \fig~\ref{subfig:dpdk-based-vhost-dsa-flow-and-diagram}. Specifically, we employ a three-stage software pipeline for each iteration of enqueue/dequeue operation. First, we check the \accel's completion record for the packet copy in the last iteration, and then write back the used descriptors using CPU core. Then, we fetch the packet descriptors from the virtqueue, assembling and submitting the corresponding \accel descriptors to \accel WQ(s) for processing. Last, while waiting for the \accel to complete the packet movement, we continue to execute other functions on the CPU cores, such as packet processing. This way, we improve the utilization of both the CPU core and \accel hardware.

The second optimization we take, aligned with \textbf{G1}, is applying batch descriptor. This is a natural fit of DPDK Vhost, as it natively applies batched (burst) processing technique to improve packet processing efficiency. For each burst of packets (typical burst size is 32) in the enqueue/dequeue operations, we prepare a single \accel batch descriptor for each iteration, which is generally enough to achieve optimal throughput.

Further, we leverage \accel's flexible group configuration feature to adopt \accel to DPDK Vhost's real usage scenarios. In a typical virtualized environment where DPDK Vhost is employed, one virtqueue may get packets from different threads (VMs or NIC queues). Also, the number of available \accel DWQ is typically smaller than the number of virtqueues. \accel's flexibility allows us to also enable multiple DWQs and let different virtqueue threads share the DWQ access for easy programming and high performance (\textbf{G6}). When multiple virtqueue threads try to submit requests to the same DWQ simultaneously, the DWQ’s spinlock will force the thread to busy-wait until acquiring the lock, causing wasting CPU time on useless activities. Hence, we also bind each DWQ to the core which uses it the most frequently to mitigate such spin-lock overhead.

Regarding the data destination in the cache/memory hierarchy (\textbf{G3}), after profiling the virtualized platform, we find the transferred packets will be generally consumed by the VM or Vhost in a timely manner, exhibiting a good temporal locality. Hence, unlike many use cases, we set the hint in \accel descriptors to keep the moved packets in the LLC to avoid extra memory traffic.

The last consideration is packet ordering. DPDK Vhost processes descriptors in the order they appear in the virtqueue. However, incoming packets may from different threads with using different \accel DWQs. To avoid packet reordering and the corresponding performance loss in the VM, we design a recording array associated with each virtqueue. This array keeps track of all on-the-fly packet copies conducted by \accel and marks the completed packet copies. When the CPU core executes step (3) of a (batched) enqueue/dequeue operation, it will scan the reordering array to find the first uncompleted copy, and write back all descriptors before it to the virtqueue, so that the VM always receives packets in order. 

Applying all these optimizations, we use the common workload DPDK-TestPMD to test the packet forwarding performance. Specifically, in host, TestPMD runs in \texttt{macfwd} mode bound with one Intel\textsuperscript{\textregistered} E810 NIC port (receiving 100Gb network traffic from a separate traffic generator) and one Vhost PMD port. In guest, TestPMD runs in \texttt{macfwd} mode bounded with a VirtIO port.  We plot the packet forwarding rate over varying packet sizes with and without \accel acceleration in \fig~\ref{subfig:dpdk-based-vhost-dsa-accel}. The fowarding rate with \accel acceleration remains constant with increasing packet sizes and for packet sizes above 256 bytes, shows a 1.14$\sim$2.29$\times$ improvement over CPU based packet forwarding.

\section{Related Work}\label{sec:related-works}

Vaidyanathan, \textit{et al.} pioneered the exploration of use cases and optimization of the efficiency of I/OAT~\cite{4629228, 4228207}. His team's prior work showcased the networking benefits and usage potential that can be gained when effectively using I/OAT. Through their findings, I/OAT proved to lower CPU utilization, improve throughput, and increase the total number of possible transactions for their tested datacenter environment. Other work of theirs highlighted the importance and improvements that are possible through well-designed, asynchronous memory copies through I/OAT \cite{4228207}. 
I/OAT has also been applied in tiered memory systems and file systems with byte-addressable persistent memory~\cite{285754,258860,10.1145/3419111.3421294,10.1145/3477132.3483550,pmem-patch}. 
\accel provides further flexibility with more user control and greater operational support compared with prior technologies.

\section{Conclusion}\label{sec:conclusion}
The new \accel opens up new opportunities for greater offloading potential through improved performance and flexibility. Compared to previous work, the additional user-side configurable options along with the ability to batch offload tasks significantly improve throughput and provide users with new ways of using accelerators on streaming data. In this paper, we also note other features supported through \accel and the accelerator's inherent benefits like mitigating cache pollution caused by co-running applications on the host processor. Through the fast-growing software ecosystem and wide range of applicable workloads, we hope this paper can motivate \accel adoption across infrastructure libraries and applications.

\bibliographystyle{plain}
\bibliography{refs}

\begin{thebibliography}{10}

\bibitem{scalable:isca:2015}
Junwhan Ahn, Sungpack Hong, Sungjoo Yoo, Onur Mutlu, and Kiyoung Choi.
\newblock A scalable processing-in-memory accelerator for parallel graph
  processing.
\newblock In {\em International Symposium on Computer Architecture, (ISCA'15)},
  2015.

\bibitem{alian2022idio}
Mohammad Alian, Siddharth Agarwal, Jongmin Shin, Neel Patel, Yifan Yuan,
  Daehoon Kim, Ren Wang, and Nam~Sung Kim.
\newblock {IDIO}: Network-driven, inbound network data orchestration on server
  processors.
\newblock In {\em 55th IEEE/ACM International Symposium on Microarchitecture,
  (MICRO'22)}, 2022.

\bibitem{alian2020data}
Mohammad Alian, Yifan Yuan, Jie Zhang, Ren Wang, Myoungsoo Jung, and Nam~Sung
  Kim.
\newblock Data direct {I/O} characterization for future {I/O} system
  exploration.
\newblock In {\em 2020 IEEE International Symposium on Performance Analysis of
  Systems and Software, (ISPASS'20)}, 2020.

\bibitem{258860}
Thomas~E. Anderson, Marco Canini, Jongyul Kim, Dejan Kosti{\'c}, Youngjin Kwon,
  Simon Peter, Waleed Reda, Henry~N. Schuh, and Emmett Witchel.
\newblock Assise: Performance and availability via client-local {NVM} in a
  distributed file system.
\newblock In {\em 14th USENIX Symposium on Operating Systems Design and
  Implementation, (OSDI'20)}, 2020.

\bibitem{flexlearn:micro:2019}
Eunjin Baek, Hunjun Lee, Youngsok Kim, and Jangwoo Kim.
\newblock {FlexLearn}: Fast and highly efficient brain simulations using
  flexible on-chip learning.
\newblock In {\em IEEE/ACM International Symposium on Microarchitecture,
  (MICRO'19)}, 2019.

\bibitem{10.5555/3488766.3488810}
Benjamin Berg, Daniel~S. Berger, Sara McAllister, Isaac Grosof, Sathya
  Gunasekar, Jimmy Lu, Michael Uhlar, Jim Carrig, Nathan Beckmann, Mor
  Harchol-Balter, and Gregory~R. Ganger.
\newblock The {CacheLib} caching engine: {Design} and experiences at scale.
\newblock In {\em Proceedings of the 14th USENIX Conference on Operating
  Systems Design and Implementation, (OSDI'20)}, 2020.

\bibitem{diannao:asplos:2014}
Tianshi Chen, Zidong Du, Ninghui Sun, Jia Wang, Chengyong Wu, Yunji Chen, and
  Olivier Temam.
\newblock {DianNao}: A small-footprint high-throughput accelerator for
  ubiquitous machine-learning.
\newblock In {\em International Conference on Architectural Support for
  Programming Languages and Operating Systems, (ASPLOS'14)}, 2014.

\bibitem{eyeriss:isca:2016}
Yu-Hsin Chen, Joel Emer, and Vivienne Sze.
\newblock Eyeriss: A spatial architecture for energy-efficient dataflow for
  convolutional neural networks.
\newblock In {\em International Symposium on Computer Architecture, (ISCA'16)},
  2016.

\bibitem{dadiannao:micro:2014}
Yunji Chen, Tao Luo, Shaoli Liu, Shijin Zhang, Liqiang He, Jia Wang, Ling Li,
  Tianshi Chen, Zhiwei Xu, Ninghui Sun, and Olivier Temam.
\newblock {DaDianNao: A Machine-Learning Supercomputer}.
\newblock In {\em IEEE/ACM International Symposium on Microarchitecture,
  (MICRO'14)}, 2014.

\bibitem{18642woo73:online}
Intel Corporation.
\newblock Dual-core {Intel Xeon Processor 5100 Series}.
\newblock
  \url{https://www.sas.com/partners/directory/intel/XeonProcessorProdBrief.pdf},
  2006.
\newblock (Accessed on 12/02/2022).

\bibitem{IntelIO3:online}
Intel Corporation.
\newblock Intel {I/O Acceleration Technology}.
\newblock
  \url{https://www.intel.com/content/www/us/en/wireless-network/accel-technology.html},
  2006.

\bibitem{WhitePap2:online}
Intel Corporation.
\newblock White paper: Accelerating high-speed networking with {Intel I/OAT}.
\newblock
  \url{https://www.intel.com/content/www/us/en/io/i-o-acceleration-technology-paper.html},
  2006.

\bibitem{quickdat48:online}
Intel Corporation.
\newblock Intel {QuickData Technology} software guide for {Linux}.
\newblock
  \url{https://www.intel.com/content/dam/doc/white-paper/quickdata-technology-software-guide-for-linux-paper.pdf},
  2008.

\bibitem{spdk:online}
Intel Corporation.
\newblock {SPDK}: Introduction to the storage performance development kit
  ({SPDK}).
\newblock
  \url{https://www.intel.com/content/www/us/en/developer/articles/tool/introduction-to-the-storage-performance-development-kit-spdk.html},
  2015.

\bibitem{GitHubin30:online}
Intel Corporation.
\newblock Github - intel/intel-cmt-cat: User space software for {Intel Resource
  Director Technology}.
\newblock \url{https://github.com/intel/intel-cmt-cat}, 2016.

\bibitem{IntelDa26:online}
Intel Corporation.
\newblock {Intel Data Direct I/O Technology}.
\newblock
  \url{https://www.intel.com/content/www/us/en/io/data-direct-i-o-technology.html},
  2018.

\bibitem{Introduc28:online}
Intel Corporation.
\newblock Introducing {Intel Scalable I/O Virtualization}.
\newblock
  \url{https://www.intel.com/content/www/us/en/developer/articles/technical/introducing-intel-scalable-io-virtualization.html},
  2018.

\bibitem{idxd-config:online}
Intel Corporation.
\newblock Github - intel/idxd-config: Utility library for controlling and
  configuring {DSA} ({Data-Streaming Accelerator}) sub-system in the {Linux}
  kernel.
\newblock \url{https://github.com/intel/idxd-config}, 2019.

\bibitem{Introduc99:online}
Intel Corporation.
\newblock Introducing the {Intel Data Streaming Accelerator (Intel DSA)} |
  01.org.
\newblock
  \url{https://01.org/blogs/2019/introducing-intel-data-streaming-accelerator},
  2019.

\bibitem{intelgna31:online}
Intel Corporation.
\newblock {GNA} - gaussian \& neural accelerator library repository.
\newblock \url{https://github.com/intel/gna}, 2022.

\bibitem{daos-dsa-accel:online}
Intel Corporation.
\newblock Performance evolution of {DAOS} servers - a sneak preview of {DAOS}
  on 4th {Gen Intel Xeon Scalable Processors}, formerly codenamed {Sapphire
  Rapids}.
\newblock
  \url{https://www.intel.com/content/www/us/en/high-performance-computing/performance-evolution-of-daos-servers.html},
  2022.

\bibitem{vtp}
Intel Corporation.
\newblock Optimized real-time video transport using {Intel Data Streaming
  Accelerator}.
\newblock
  \url{https://networkbuilders.intel.com/solutionslibrary/optimized-real-time-video-transport-using-intel-data-streaming-accelerator},
  2023.

\bibitem{intel-agi}
Intel Corporation.
\newblock {Intel® Agilex™ 7 FPGA I-Series Development Kit }.
\newblock
  \\\url{https://www.intel.com/content/www/us/en/products/details/fpga/development-kits/agilex/i-series/dev-agi027.html},
  accessed in 2023.

\bibitem{dml}
Intel Corporation.
\newblock {Intel® Data Mover Library (Intel® DML)}.
\newblock \\\url{https://github.com/intel/DML}, accessed in 2023.

\bibitem{dsa-spec}
Intel Corporation.
\newblock {Intel® Data Streaming Accelerator Architecture Specification}.
\newblock
  \\\url{https://software.intel.com/en-us/download/intel-data-streaming-accelerator-preliminary-architecture-specification},
  accessed in 2023.

\bibitem{isa-l}
Intel Corporation.
\newblock {Intel® Intelligent Storage Acceleration Library (Intel® ISA-L)}.
\newblock \\\url{https://github.com/intel/isa-l}, accessed in 2023.

\bibitem{pcm}
Intel Corporation.
\newblock {Intel® Performance Counter Monitor (Intel® PCM)}.
\newblock \\\url{https://github.com/intel/pcm}, accessed in 2023.

\bibitem{cxl1}
{CXL Consortium}.
\newblock {Compute Express Link (CXL)}.
\newblock \\\url{https://www.computeexpresslink.org}, accessed in 2021.

\bibitem{towardsgen:micro:2019}
Vidushi Dadu, Jian Weng, Sihao Liu, and Tony Nowatzki.
\newblock Towards general purpose acceleration by exploiting common
  data-dependence forms.
\newblock In {\em 51st IEEE/ACM International Symposium on Microarchitecture,
  (MICRO'19)}, 2019.

\bibitem{daos:online}
DAOS.
\newblock Distributed asynchronous object storage ({DAOS}).
\newblock
  \url{https://www.intel.com/content/www/us/en/high-performance-computing/daos.html},
  2019.

\bibitem{de2010new}
Arnaldo~Carvalho De~Melo.
\newblock The new {Linux} "perf" tools.
\newblock In {\em Slides from Linux Kongress}, volume~18, pages 1--42, 2010.

\bibitem{darksilicon:isca:2011}
Hadi Esmaeilzadeh, Emily Blem, Renee St.~Amant, Karthikeyan Sankaralingam, and
  Doug Burger.
\newblock Dark silicon and the end of multicore scaling.
\newblock In {\em Proceedings of the 38th Annual International Symposium on
  Computer Architecture, (ISCA'11)}, 2011.

\bibitem{cachebench:online}
Facebook.
\newblock Cachebench: Benchmark and stress testing tool to evaluate cache
  performance with real hardware and real cache workloads.
\newblock
  \url{https://cachelib.org/docs/Cache_Library_User_Guides/Cachebench_Overview},
  2020.

\bibitem{goglin2008improving}
Brice Goglin.
\newblock Improving message passing over {Ethernet} with {I/OAT} copy offload
  in open-mx.
\newblock In {\em 2008 IEEE International Conference on Cluster Computing,
  (IEEE Cluster'08)}, 2008.

\bibitem{xmem}
Mark Gottscho, Sriram Govindan, Bikash Sharma, Mohammed Shoaib, and Puneet
  Gupta.
\newblock X-mem: A cross-platform and extensible memory characterization tool
  for the cloud.
\newblock In {\em 2016 IEEE International Symposium on Performance Analysis of
  Systems and Software, (ISPASS'16)}, 2016.

\bibitem{7312664}
Paul Grun, Sean Hefty, Sayantan Sur, David Goodell, Robert~D. Russell, Howard
  Pritchard, and Jeffrey~M. Squyres.
\newblock A brief introduction to the openfabrics interfaces - a new network
  {API} for maximizing high performance application efficiency.
\newblock In {\em 2015 IEEE 23rd Annual Symposium on High-Performance
  Interconnects}, 2015.

\bibitem{graphicionado:micro:2016}
Tae~Jun Ham, Lisa Wu, Narayanan Sundaram, Nadathur Satish, and Margaret
  Martonosi.
\newblock Graphicionado: A high-performance and energy-efficient accelerator
  for graph analytics.
\newblock In {\em IEEE/ACM International Symposium on Microarchitecture,
  (MICRO'16)}, 2016.

\bibitem{extensor:micro:2019}
Kartik Hegde, Hadi Asghari-Moghaddam, Michael Pellauer, Neal Crago, Aamer
  Jaleel, Edgar Solomonik, Joel Emer, and Christopher~W Fletcher.
\newblock Extensor: An accelerator for sparse tensor algebra.
\newblock In {\em IEEE/ACM International Symposium on Microarchitecture,
  (MICRO'19)}, 2019.

\bibitem{huggahalli2005direct}
Ram Huggahalli, Ravi Iyer, and Scott Tetrick.
\newblock Direct cache access for high bandwidth network {I/O}.
\newblock In {\em 32nd International Symposium on Computer Architecture,
  (ISCA'05)}, 2005.

\bibitem{TPU}
Norman~P. Jouppi, Cliff Young, Nishant Patil, David Patterson, Gaurav Agrawal,
  Raminder Bajwa, Sarah Bates, Suresh Bhatia, Nan Boden, Al~Borchers, Rick
  Boyle, Pierre-luc Cantin, Clifford Chao, Chris Clark, Jeremy Coriell, Mike
  Daley, Matt Dau, Jeffrey Dean, Ben Gelb, Tara~Vazir Ghaemmaghami, Rajendra
  Gottipati, William Gulland, Robert Hagmann, C.~Richard Ho, Doug Hogberg, John
  Hu, Robert Hundt, Dan Hurt, Julian Ibarz, Aaron Jaffey, Alek Jaworski,
  Alexander Kaplan, Harshit Khaitan, Daniel Killebrew, Andy Koch, Naveen Kumar,
  Steve Lacy, James Laudon, James Law, Diemthu Le, Chris Leary, Zhuyuan Liu,
  Kyle Lucke, Alan Lundin, Gordon MacKean, Adriana Maggiore, Maire Mahony,
  Kieran Miller, Rahul Nagarajan, Ravi Narayanaswami, Ray Ni, Kathy Nix, Thomas
  Norrie, Mark Omernick, Narayana Penukonda, Andy Phelps, Jonathan Ross, Matt
  Ross, Amir Salek, Emad Samadiani, Chris Severn, Gregory Sizikov, Matthew
  Snelham, Jed Souter, Dan Steinberg, Andy Swing, Mercedes Tan, Gregory
  Thorson, Bo~Tian, Horia Toma, Erick Tuttle, Vijay Vasudevan, Richard Walter,
  Walter Wang, Eric Wilcox, and Doe~Hyun Yoon.
\newblock In-datacenter performance analysis of a {Tensor Processing Unit}.
\newblock In {\em 44th ACM/IEEE International Symposium on Computer
  Architecture, (ISCA'17)}, 2017.

\bibitem{10.1145/3419111.3421294}
Anuj Kalia, David Andersen, and Michael Kaminsky.
\newblock Challenges and solutions for fast remote persistent memory access.
\newblock In {\em 11th ACM Symposium on Cloud Computing, (SoCC'20)}, 2020.

\bibitem{Kanev:2015:PWC:2749469.2750392}
Svilen Kanev, Juan~Pablo Darago, Kim Hazelwood, Parthasarathy Ranganathan, Tipp
  Moseley, Gu-Yeon Wei, and David Brooks.
\newblock Profiling a warehouse-scale computer.
\newblock In {\em 42nd IEEE/ACM International Symposium on Computer
  Architecture, (ISCA'15)}, 2015.

\bibitem{proto:micro:2021}
Sagar Karandikar, Chris Leary, Chris Kennelly, Jerry Zhao, Dinesh Parimi,
  Borivoje Nikolic, Krste Asanovic, and Parthasarathy Ranganathan.
\newblock A hardware accelerator for protocol buffers.
\newblock In {\em 54th IEEE/ACM International Symposium on Microarchitecture,
  (MICRO'21)}, 2021.

\bibitem{osu-micro}
Network-Based~Computing Laboratory.
\newblock Osu micro-benchmarks 7.0.
\newblock \url{http://mvapich.cse.ohio-state.edu/benchmarks/}, 2022.
\newblock (Accessed on 01/28/2023).

\bibitem{masterofnone:isca:2019}
Andrea Lottarini, Jo\~{a}o~P. Cerqueira, Thomas~J. Repetti, Stephen~A. Edwards,
  Kenneth~A. Ross, Mingoo Seok, and Martha~A. Kim.
\newblock Master of none acceleration: A comparison of accelerator
  architectures for analytical query processing.
\newblock In {\em 47th ACM/IEEE International Symposium on Computer
  Architecture, (ISCA'19)}, 2019.

\bibitem{tabla:hpca:2016}
Divya Mahajan, Jongse Park, Emmanuel Amaro, Hardik Sharma, Amir Yazdanbakhsh,
  Joon Kim, and Hadi Esmaeilzadeh.
\newblock {TABLA}: A unified template-based framework for accelerating
  statistical machine learning.
\newblock In {\em IEEE International Symposium on High-Performance Computer
  Architecture, (HPCA'16)}, 2016.

\bibitem{mlperf}
MLCommons.
\newblock Mlperf benchmark.
\newblock \\\url{https://mlcommons.org/en/training-normal-10/}, 2022.
\newblock (Accessed on 01/28/2023).

\bibitem{streamdf:isca:2017}
Tony Nowatzki, Vinay Gangadhar, Newsha Ardalani, and Karthikeyan Sankaralingam.
\newblock {Stream-dataflow Acceleration}.
\newblock In {\em International Symposium on Computer Architecture, (ISCA'17)},
  2017.

\bibitem{vpp}
The Fast~Data Project.
\newblock Fd.io -- the world’s secure networking data plane.
\newblock \url{https://fd.io}.

\bibitem{dsa-perf-micros:online}
Nikhil Rao, Nirav Shah, and Michael Beale.
\newblock Github - intel/dsa-perf-micros: Intel® dsa performance micros.
\newblock \url{https://github.com/intel/dsa-perf-micros}, 12 2022.
\newblock (Accessed on 1/05/2023).

\bibitem{10.1145/3477132.3483550}
Amanda Raybuck, Tim Stamler, Wei Zhang, Mattan Erez, and Simon Peter.
\newblock {HeMem: Scalable} tiered memory management for big data applications
  and real {NVM}.
\newblock In {\em ACM SIGOPS 28th Symposium on Operating Systems Principles
  (SOSP'21)}, 2021.

\bibitem{cxl2}
Debendra~Das Sharma.
\newblock Compute express linktm (cxltm): Enabling heterogeneous data-centric
  computing with heterogeneous memory hierarchy.
\newblock {\em IEEE Micro}, pages 1--7, 2022.

\bibitem{dnnweaver:micro:2016}
Hardik Sharma, Jongse Park, Divya Mahajan, Emmanuel Amaro, Joon Kim, Chenkai
  Shao, Asit Misra, and Hadi Esmaeilzadeh.
\newblock From high-level deep neural models to {FPGA}s.
\newblock In {\em 49th IEEE/ACM International Symposium on Microarchitecture,
  (MICRO'16)}, 2016.

\bibitem{285754}
Jingbo Su, Jiahao Li, Luofan Chen, Cheng Li, Kai Zhang, Liang Yang, and Yinlong
  Xu.
\newblock Revitalizing the forgotten on-chip {DMA} to expedite data movement in
  {NVM-based} storage systems.
\newblock In {\em Proceedings of the 21st USENIX Conference on File and Storage
  Technologies, (FAST'23)}, 2023.

\bibitem{10.1145/3555050.3569118}
Qiang Su, Chuanwen Wang, Zhixiong Niu, Ran Shu, Peng Cheng, Yongqiang Xiong,
  Dongsu Han, Chun~Jason Xue, and Hong Xu.
\newblock {PipeDevice}: {A} hardware-software co-design approach to intra-host
  container communication.
\newblock In {\em Proceedings of the 18th International Conference on Emerging
  Networking EXperiments and Technologies, (CoNEXT'22)}, 2022.

\bibitem{ibm-dma}
Yutaka Sugawara, Dong Chen, Ruud~A. Haring, Abdullah Kayi, Eugene Ratzlaff,
  Robert~M. Senger, Krishnan Sugavanam, Ralph Bellofatto, Ben~J. Nathanson, and
  Craig Stunkel.
\newblock Data movement accelerator engines on a prototype {Power10} processor.
\newblock {\em IEEE Micro}, 2022.

\bibitem{tootoonchian2018resq}
Amin Tootoonchian, Aurojit Panda, Chang Lan, Melvin Walls, Katerina Argyraki,
  Sylvia Ratnasamy, and Scott Shenker.
\newblock {ResQ}: Enabling {SLO}s in network function virtualization.
\newblock In {\em 15th USENIX Symposium on Networked Systems Design and
  Implementation, (NSDI'18)}, 2018.

\bibitem{4629228}
K.~Vaidyanathan, L.~Chai, W.~Huang, and D.~K. Panda.
\newblock Efficient asynchronous memory copy operations on multi-core systems
  and {I/OAT}.
\newblock In {\em 2007 IEEE International Conference on Cluster Computing,
  (IEEE Cluster'07)}, 2007.

\bibitem{4228207}
K.~Vaidyanathan, W.~Huang, L.~Chai, and D.~K. Panda.
\newblock Designing efficient asynchronous memory operations using hardware
  copy engine: A case study with {I/OAT}.
\newblock In {\em 2007 IEEE International Parallel and Distributed Processing
  Symposium, (IPDPS'07)}, 2007.

\bibitem{vaidyanathan2007benefits}
Karthikeyan Vaidyanathan and Dhabaleswar~K Panda.
\newblock Benefits of {I/O} acceleration technology ({I/OAT}) in clusters.
\newblock In {\em 2007 IEEE International Symposium on Performance Analysis of
  Systems \& Software, (ISPASS'07)}, 2007.

\bibitem{iouring}
WikiPedia.
\newblock io\_uring.
\newblock \url{https://en.wikipedia.org/wiki/Io\_uring}, 2022.

\bibitem{pmem-patch}
Zi~Yan.
\newblock Accelerate page migration and use memcg for {PMEM} management.
\newblock \\\url{https://lwn.net/Articles/784925/}, accessed in 2023.

\bibitem{yazdanbakhsh2021evaluation}
Amir Yazdanbakhsh, Kiran Seshadri, Berkin Akin, James Laudon, and Ravi
  Narayanaswami.
\newblock An evaluation of edge {TPU} accelerators for convolutional neural
  networks.
\newblock {\em arXiv preprint arXiv:2102.10423}, 2021.

\bibitem{yuan2021don}
Yifan Yuan, Mohammad Alian, Yipeng Wang, Ren Wang, Ilia Kurakin, Charlie Tai,
  and Nam~Sung Kim.
\newblock Don’t forget the {I/O} when allocating your {LLC}.
\newblock In {\em 2021 ACM/IEEE 48th Annual International Symposium on Computer
  Architecture, (ISCA'21)}, 2021.

\bibitem{ntb-driver}
The Kernel Development Community.
\newblock Non-Transparent Bridge Drivers
\newblock \\\url{https://www.kernel.org/doc/html/latest/driver-api/ntb.html}, accessed in 2023.
  
\bibitem{dpdk-vhost-lib:online}
DPDK
\newblock DPDK: Vhost library
\newblock \\\url{https://doc.dpdk.org/guides/prog_guide/vhost_lib.html}, accessed in 2023.

\end{thebibliography}
\clearpage

\appendix

In this appendix, we discuss three real-world applications that benefit from \accel enablement. Through active community development, software stacks for data-intensive scenarios (e.g., networking, storage, caching, ML/HPC) have had proper support for \accel -- paving the way for related workload acceleration.

\section{Libfabric in HPC/ML}
\accel can also be applied in a wide range of applications. One example is machine learning (ML) processing and HPC frameworks. On the one hand, as an iterative process, machine learning processing requires clearing certain memory regions (e.g., gradient vectors) before it can proceed to the next iteration; on the other hand, in the scenarios of distributed ML, a huge amount of tensors and vectors need to be copied and transferred across nodes in a cluster via communication libraries like \texttt{libfabric}~\cite{7312664}. As ML model sizes increase rapidly over the years, such operations can create a significant burden to the CPU. With \accel's assistance, such memory-zeroing and memory-copy routines can be offloaded and the CPU cycles can be saved for more critical tasks. Note that similar problems also occur in certain HPC workloads, where \accel can be leveraged. 
In this work, we use three workloads on top of \texttt{libfabric} which offload \texttt{memcpy} to \accel as an example to demonstrate its benefits. \texttt{libfabric} is a framework exporting fabric communication services to applications with low-level communication APIs. Note that we use Segmentation and Reassembly (SAR) protocol for larger messages when Linux Cross Memory Attach (CMA) is not permitted.

\begin{figure}[!t]
    \centering
    \subfloat[Libfabric microbenchmark throughputs\label{subfig:libfabric-throughputs}]{%
        \includegraphics[width=0.49\textwidth]{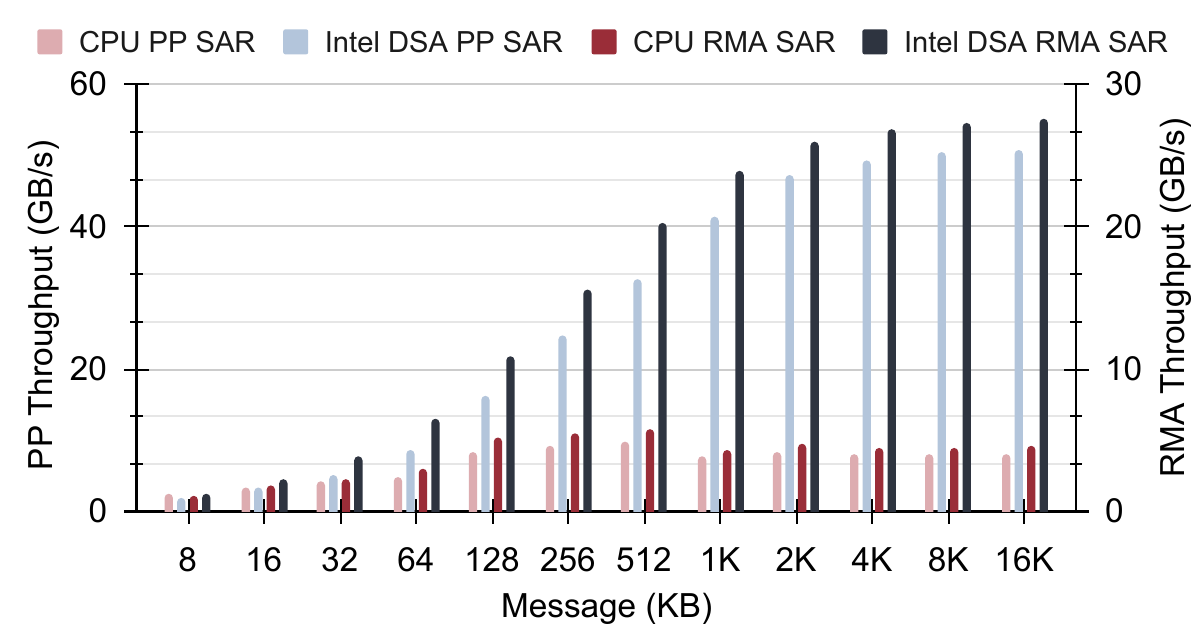}%
    }
    \hfill
    \subfloat[OSU's MPI microbenchmark improvements\label{subfig:osu-mpi-improvements}]{%
        \includegraphics[width=0.49\textwidth]{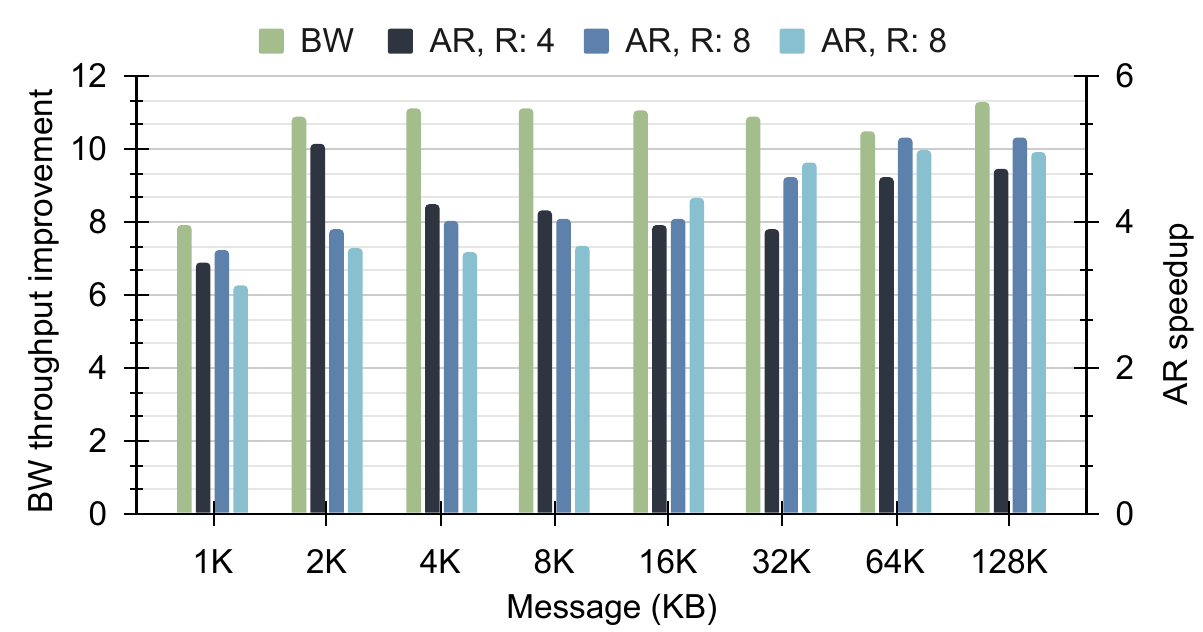}%
    }
    \caption{\texttt{libfabric}-based experiment results with CPU and \accel.}
    \label{fig:libfabric}
\end{figure}

The first workload is \texttt{libfabric}'s native microbenchmark for throughput measurement. It has two example uses of operation: Pingpong (PP) where two endpoints exchange messages in a pingpong pattern, and remote memory access (RMA) where the remote memory read/write bandwidth is tested between two endpoints. As shown in \fig~\ref{subfig:libfabric-throughputs} when message size increases (32KB or larger), \accel gradually outperforms CPU by as high as 5.1$\times$ for the Pingpong test. The same trend happens for the RMA test, where \accel can enjoy as high as a 4.7$\times$ speedup.

The second workload is OSU's MPI benchmark~\cite{osu-micro}. We test two common use cases: one-direction bandwidth test (BW) between two endpoints, and MPI\_AllReduce collective (AR) with two or eight ranks (R). \fig~\ref{subfig:osu-mpi-improvements} compares the results between software and \accel. Similar to the first workload, we find that \accel has a significant advantage over CPU solutions, especially when the message size is larger than 1MB, which can be 5.1$\times$, 5.2$\times$, 5.0$\times$, for 2-, 4-, and 8-rank settings respectively. 

We further run a PyTorch-based MLPerf~\cite{mlperf} BERT pretraining workload on top of MPI\_AllReduce with two or eight ranks (R).  
We observed that, as a large language model, BERT needs to transfer a huge amount of data per iteration, giving \accel opportunities for acceleration -- 2.8$\times$ AR speedup in the 2-rank setting compared to software, and the speedup will further increase to 3.3$\times$ for the 8-rank setting. This also leads to a 3.7\% and 8.8\% speedup for end-to-end training time respectively. 

\begin{figure}
    \centering
    \centerline{\includegraphics[width=0.9\columnwidth]{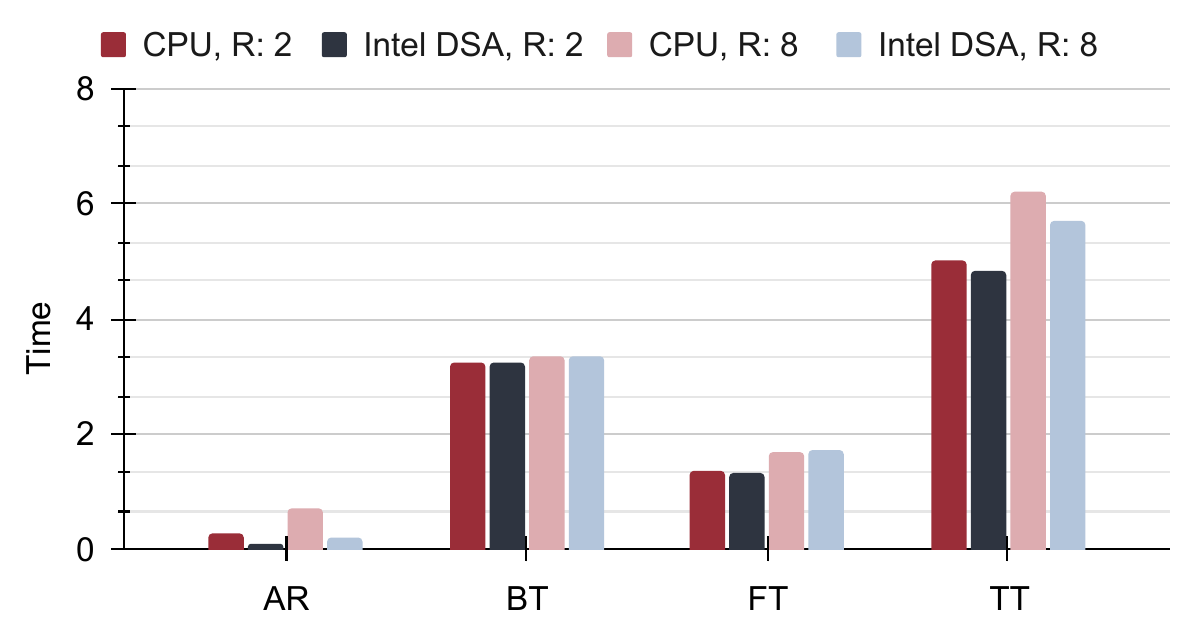}}
    \vspace{-6pt}
    \caption{BERT phase timings}
    \label{fig:bert-timings}
\end{figure}

\begin{figure*}[!t]
    \centering
    \subfloat[Get \& Set relative rate\label{subfig:cachebench-op-rate}]{%
        \includegraphics[width=0.49\textwidth]{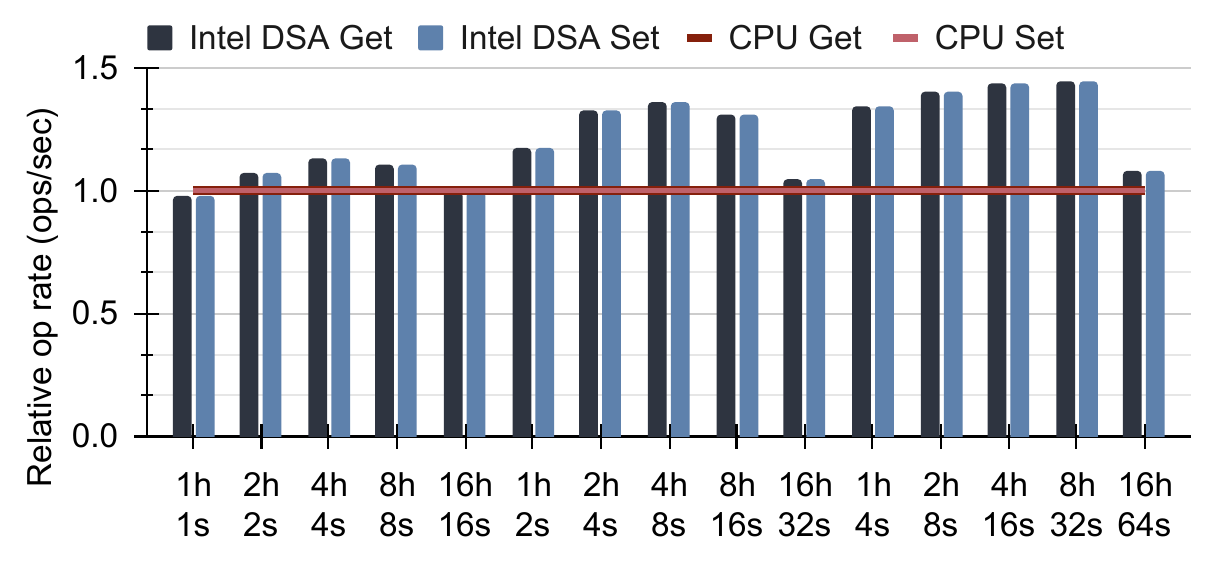}%
    }
    \hfill
    \subfloat[Find \& Alloc 99.999\% tail latency\label{subfig:cachebench-tail-latency}]{%
        \includegraphics[width=0.49\textwidth]{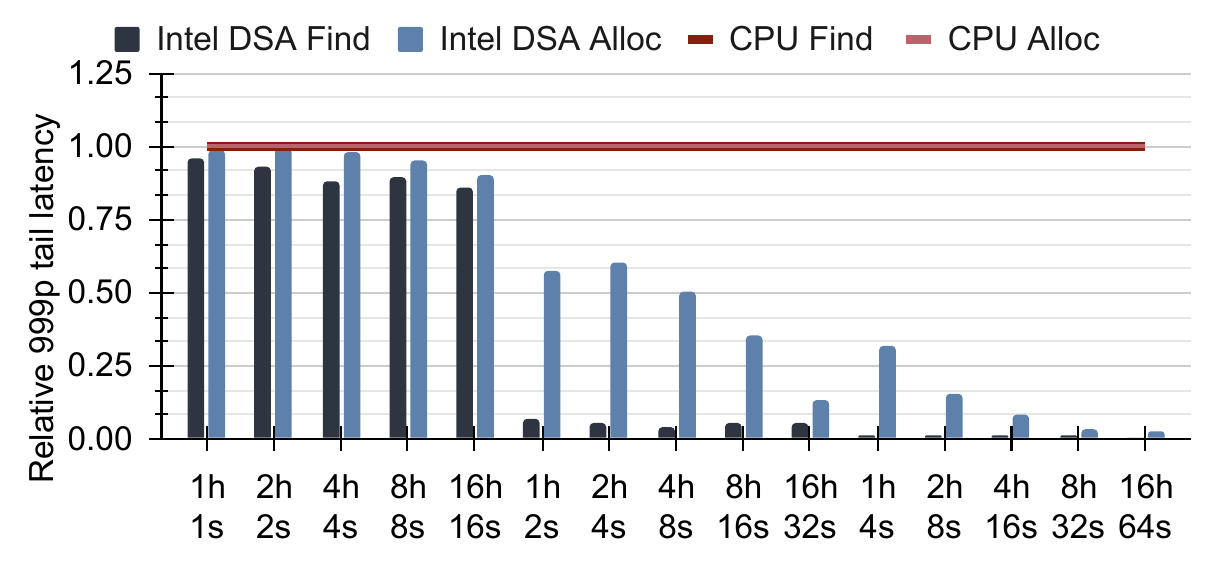}%
    }
    \caption{CacheBench relative operational throughput and tail latency improvements for their corresponding operations when offloading through \accel. Configurations list the number of hardware cores (\#h) and the number of software threads (\#s). Testing was done using four shared \accel work queues.}
    \label{fig:cachebench}
\end{figure*}

\section{CacheLib-based Cloud Data Caching Service} \label{subsec:cachelib}
Today's online cloud web services frequently employ caching systems to improve system performance such as response time and efficiency. Caching systems typically involve intensive data operations for data manipulation, creating a significant burden on the cores. In this section, we use Meta's state-of-the-art open-source caching engine -- CacheLib -- to demonstrate how \accel can help alleviate such overheads. CacheLib~\cite{10.5555/3488766.3488810} is a performant and versatile library providing common software caching functionalities. We used CacheBench ~\cite{cachebench:online} to drive CacheLib for measuring cache effectiveness and performance in a simulated cloud environment. Through this benchmark, get and set operations are performed on a pre-configured cache for evaluating operation throughput and latency characteristics given \texttt{find()} (i.e. get) and \texttt{allocate()} (i.e. set) API calls. Through our tested configuration, we used a 64GB cache and ran two million get operations per thread. 

The get and set operations involve frequent memory copies of various transfer sizes via \memcpy, and thus, open opportunities for \accel offloading. Since \accel improves throughput and performance for memory copies generally at or above 8KB, we offload any \memcpys above this threshold. From our testing, around 4.8\% of \memcpys are copying data of 8KB or larger in size, but accounts for 96.4\% of data copied.

To achieve offloading to \accel without substantially modifying CacheBench, we developed \accel Transparent Offload library (DTO). Applications that wish to use the DTO library, a user can either dynamically link the libarary through using the linker options \texttt{-ldto} and \texttt{-laccel-config}, or preload the library via \texttt{LD\_PRELOAD} without having to recompile their application. When common system API calls like \memmove, \memcpy, \memset, or \memcmp are used, DTO functions by intercepting and replacing them with corresponding synchronous \accel operations. Regarding our use with CacheBench, the core  would redo offloaded operations when encountering page faults during \accel offloading.

The resulting rate of get and set operations when transparently offloading to \accel is greatly improved as seen in \fig\ref{subfig:cachebench-op-rate}. The decreased rate improvement when using more than eight cores is likely due to only having four work queues available for descriptor submission. Since these operations are offloaded synchronously, a thread must stall when all \accel groups are actively managing a descriptor. The 99.999\% tail latency additionally saw significant improvements as the performance of using large memory copy operations on \accel greatly exceeds that of relying on the cores (\fig~\ref{subfig:cachebench-tail-latency}).

\section{SPDK NVMe/TCP Target}

\begin{figure}[!t]
    \centerline{\includegraphics[width=\columnwidth]{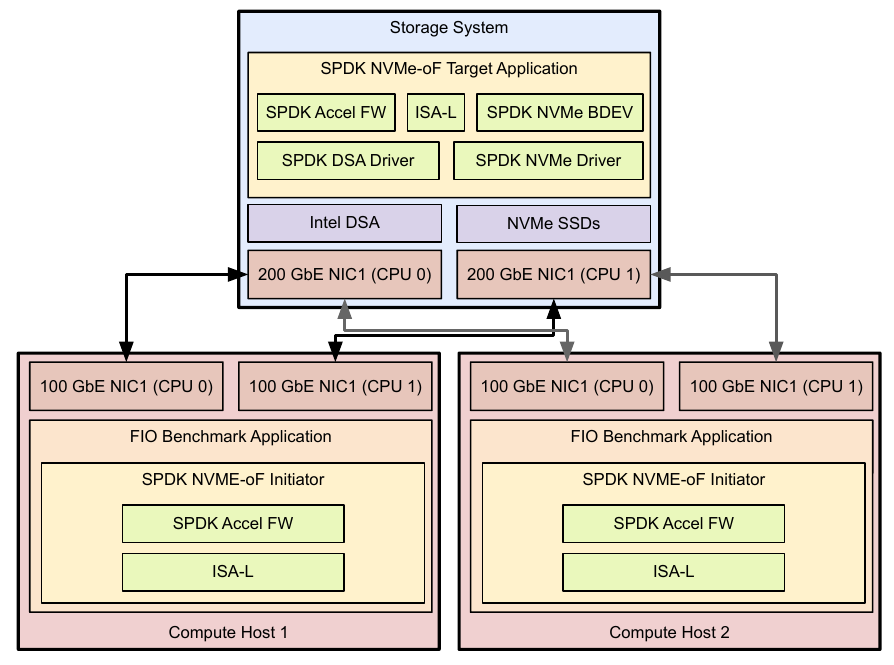}}
    \caption{Design overview of SPDK NVMe/TCP Target}
    \label{fig:spdk-based-fio-setup}
\end{figure}

SPDK contains a set of tools and libraries for designing high-performance, scalable, user-mode storage applications. SPDK achieves high performance by avoiding locking on the I/O path, hoisting the needed drivers from kernel-space to user-space, and polling for completions instead of using interrupts -- eliminating overhead costs such as context switching between the interrupt handler and the application~\cite{spdk:online}. 

\begin{figure*}[!t]
    \centering
    \subfloat[16~KB random reads\label{subfig:sdpk-16kb-rand}]{%
        \includegraphics[width=0.49\textwidth]{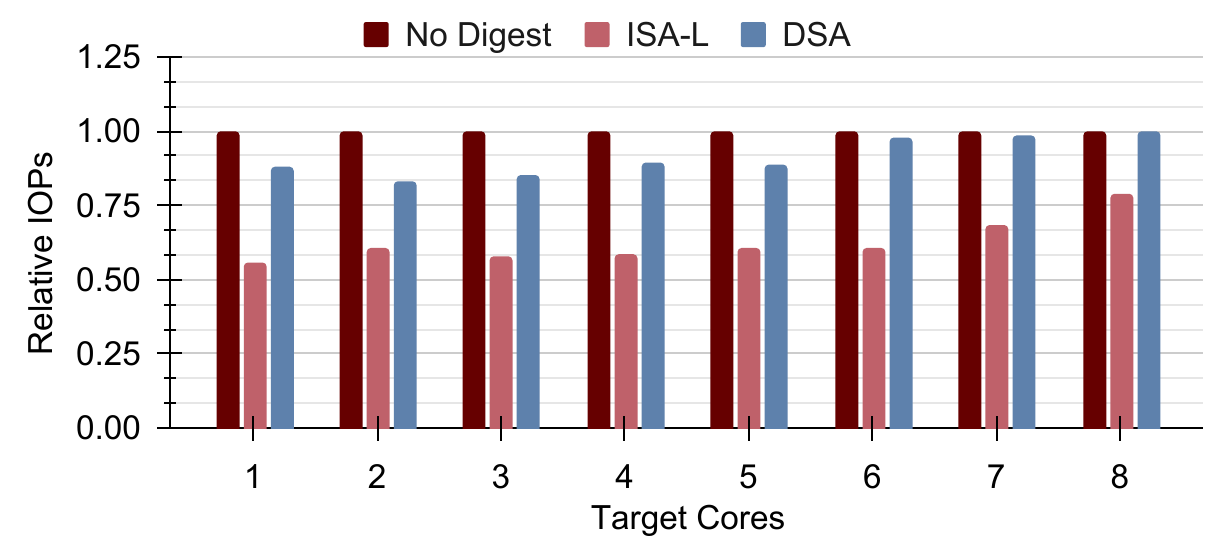}%
    }
    \hfill
    \subfloat[128~KB sequential reads\label{subfig:spdk-128kb-seq}]{%
        \includegraphics[width=0.49\textwidth]{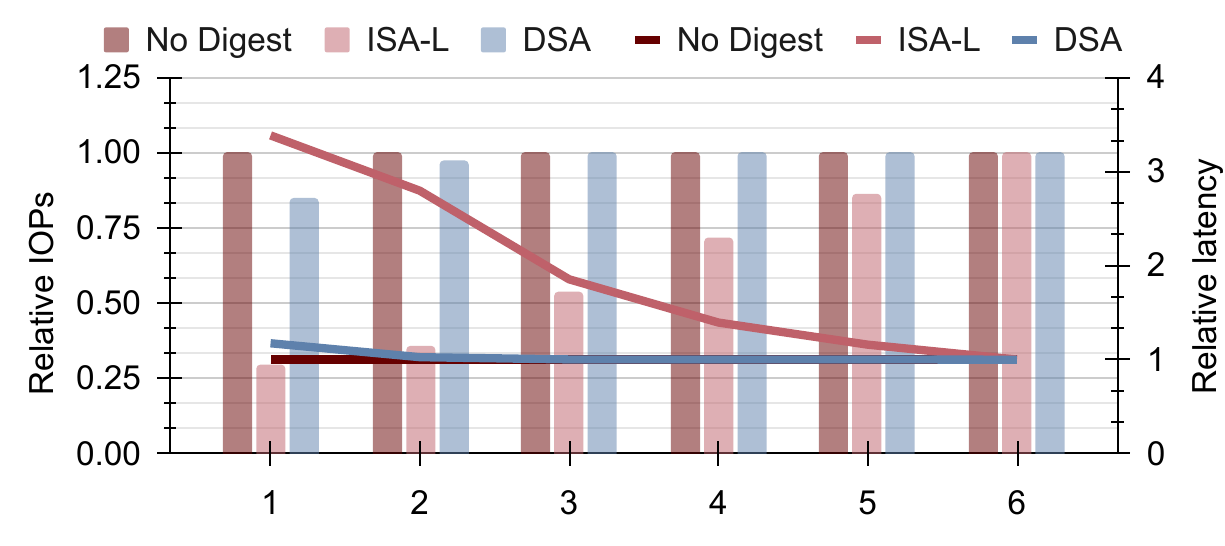}%
    }
    \caption{SPDK NVMe/TCP target with \accel CRC32 offload read IOPs}
    \label{fig:spdk-based-fio-iops-lat}
\end{figure*}

SPDK provides a high-performance NVMe/TCP target application capable of serving disks over the network or to other processes as shown in \fig~\ref{fig:spdk-based-fio-setup}. 
Protocol Data Units (PDUs) are sent between the host and controllers in the NVM subsystem. With the optional Data Digest field included in the PDUs, a CRC32 checksum is generated for strong error detection between the two TCP endpoints (i.e., target application and processes consuming the data over the network). This CRC32 checksum can be offloaded to \accel  through SPDK's acceleration framework on supported systems. Alternatively, this computation is done through SPDK's ISA-L optimized software library that uses AVX-512 vectorized instructions to improve upon the baseline CRC32 computation if available. We used the FIO benchmarking application with the data digest field enabled on the compute host to submit I/O read requests of different sizes to the SPDK NVMe/TCP target application.

Through a typical FIO TCP read workflow, an initiator system first generates a read request to a target system and transmits this request over TCP. Once received, the target system generates a CRC32 checksum, produces a TCP PDU with the checksum assigned to the Data Digest field, and sends the packet to the requesting initiator. The initiator system recomputes the checksum and confirms I/O completion when the two checksums match.

In testing, we used two ICX Xeon systems as NVMe--over--fabric (NVMe--oF) initiators and one \accel--enabled SPR system as an NVMe--oF target (\fig~\ref{fig:spdk-based-fio-setup}). FIO was used by the initiators for read workloads and communicated with the NVMe--managed target system. 16 NVMe SSDs were attached to the target system and used as general block devices. All systems used the underlying POSIX sockets transport layer API for TCP communication. For CRC32 offloading, requests are batched when possible and sent by the target to available \accel devices and polled in user-space. Specifically, work queue entry descriptors are generated by SPDK's accel framework, submitted to \accel, and polled for completion assuming device availability. If the device is either unavailable or not enabled, the framework falls back to the ISA-L CRC32 function. In this setup, the initiators must use ISA-L for checksum computation due to having no \accel devices.

\fig~\ref{subfig:spdk-128kb-seq} demonstrates the nearly equivalent average latency when using \accel for CRC32 offloads compared to when no CRC32 Data Digests are generated, and substantially lower than using ISA-L for computation. Similar trends are seen with 16~KB random read and 128~KB sequential read results in \fig\ref{fig:spdk-based-fio-iops-lat}, and importantly, both the test with no Data Digest and using \accel reach throughput saturation using only six target cores for 16~KB random read and two target cores for 128~KB sequential read while using ISA-L saturates at over 8 and 6 cores respectively.

\end{document}